\documentclass[useAMS,usenatbib]{mn2e}

\usepackage{epsfig}
\usepackage{longtable}
\usepackage{times}

\usepackage{amsmath}

\def \crab {Crab}
\def \vela {Vela\,X--1}
\def \foru {4U\,1700--377}
\def \igr1739 {XTE\,J1739--302}
\def \igr17544 {IGR\,J17544--2619}
\def \sax {SAX\,J1818.6--1703}
\def \igr16418 {IGR\,J16418--4532}
\def \igr16479 {IGR\,J16479--4514}
\def \igr16465 {IGR\,J16465--4507}
\def \h1907 {H\,1907+097}
\def \igr18483 {IGR\,J18483--0311}

\def \inte {\textit{INTEGRAL}}

\def \degmark {^\circ}

\def \ergsec{\hbox{erg s$^{-1}$}}
\def \countsec{\hbox{counts s$^{-1}$}}

\def \hcm {\hbox {\ifmmode $ atom cm$^{-2}\else atom cm$^{-2}$\fi}}

\def \ATel {Astron.\ Tel.}
\def \apj {ApJ}
\def \apjl {ApJL}

\def \aap {A\&A}

\def \pasj {PASJ}

\def \mnras {MNRAS}

\def \ssr {Space Science Reviews}
\def \ima {\textsl{IMA}}
\def \lcr {\textsl{LCR}}

\newcommand{\be}{\begin{equation}}

\newcommand{\ee}{\end{equation}}

\title[Cumulative luminosity distributions of SFXTs]{Cumulative luminosity distributions of Supergiant Fast X-ray Transients in hard X--rays}
\author[Paizis \& Sidoli]{A. Paizis$^{1}$\thanks{E-mail: ada@iasf-milano.inaf.it} \& L. Sidoli$^{1}$  \\
$^{1}$INAF, Istituto di Astrofisica Spaziale e Fisica Cosmica, Via E.\ Bassini 15,   I-20133 Milano,  Italy }

\begin{document}

\date{Accepted: 2014 January 24  Received: 2013 November 15 }

\pagerange{\pageref{firstpage}--\pageref{lastpage}} \pubyear{2012}

\maketitle

\label{firstpage}

\begin{abstract}
We have analyzed in a systematic way about nine years of \inte~data (17--100\,keV) focusing on Supergiant Fast X-ray Transients (SFXTs) and three classical High Mass X-ray Binaries (HMXBs). Our approach has been twofold: image based analysis,  sampled over a $\sim$ks time frame to investigate the long-term properties of the sources, and lightcurve based analysis,  sampled over a 100\,s time frame to seize the fast variability of each source during its $\sim$\,ks activity.  \\
We find that while the prototypical SFXTs (IGR\,J17544--2619, XTE\,J1739--302 and SAX\,J1818.6--1703)  are among the sources with the lowest $\sim$\,ks~based duty cycle ($<$1\% activity over nine years of data), when studied at the 100\,s level, they are the ones with the highest detection percentage, meaning that, when active, they tend to have many bright short-term flares with respect to the other SFXTs.\\
To investigate in a coherent and self consistent way all the available results within a physical scenario, we have extracted cumulative luminosity distributions for all the sources of the sample. The characterization of such distributions in hard X-rays, presented for the first time in this work for the SFXTs, shows that a power-law model is a plausible representation for SFXTs, while it can only reproduce the very high luminosity tail of the classical HMXBs, and even then, with a significantly steeper power-law slope with respect to SFXTs. The physical implications of these results within the frame of accretion in wind-fed systems are discussed.

\end{abstract}


        \section{Introduction\label{intro}}

\begin{table*}
 \centering
  \caption{The sources studied in this work, together with their main characteristics.}
  \begin{tabular}{@{}llllll@{}}
\hline
   Name & Companion & Distance & Orbital Period &   Spin Period &  Super-Orbital Period \\
    &  &  kpc & days   &   s &  days \\
\hline

Prototypical SFXTs &          &          &           &    & \\
\hline
XTE~J1739--302   & O8Iab(f) (1,2)   & 2.7 (2)      & 51.47 $\pm$ 0.02 (3)    &                       &  \\
IGR~J17544--2619 & O9Ib (4)         & 3.6 (2,4)    & 4.926 $\pm$ 0.001 (5)   &  71.49$\pm{0.02}$ (6) & \\
SAX~J1818.6--1703 & $\sim$B0I (7,8) & 2, 2.1 (8,9) & 30.0 $\pm$ 0.2 (10,11)  &                       &  \\%
\hline

Intermediate  SFXTs &          &          &         &      &   \\
\hline
IGR~J16418--4532 &  OB Sg               &  13   (12)      &  3.753$\pm{0.004}$ (13)     &  1212 $\pm{6}$  (14)        &   14.6842 $\pm{0.0008}$ (15, 16) \\
IGR~J16479--4514 & O8.5I, O9.5Iab (2,17)& 4.9, 2.8 (2,17) & 3.3194 $\pm$ 0.001 (18,19)  &                             &   11.880 $\pm{ 0.002}$ (15, 16) \\
IGR~J18483--0311 &  B0.5Ia  (20)        & 3 (20)          &  18.55 $\pm{0.03}$ (21)     &  21.0526$\pm{0.0005}$ (22)  &   \\
IGR~J18450--0435 & O9.5I (23)           & 3.6 (23)        &  5.7195$\pm{0.0007}$ (24)   &                             &  \\
\hline

Less explored SFXTs      &           &           &           &         &  \\
\hline
IGR~J16465--4507  & O9.5Ia  (23)       &   9.5 $^{+14.1} _{-5.7}$ (23)  &  30.243$\pm{0.035}$ (25,26)  & 228$\pm{6}$  (27)  &    \\
IGR~J08408--4503  & O8.5Ib  (28)       &   2.7 (29)                     &                              &                    &    \\
IGR~J18410--0535 &  B1Ib  (17)         &   3.2 $^{+2.0} _{-1.5}$ (23)   &                              &                   &    \\
\hline

The periodic SFXT      &           &           &           &         &  \\
\hline
IGR~J11215--5952  &  B0.5Ia   (30, 31)    &  6.2, 8   (32, 30)   &  164.6 (34,35)   &  186.78$\pm{0.3}$ (36)  &  \\
\hline
\hline 
 HMXBs  &          &          &         &     &    \\
\hline
4U~1700--377 & O6.5Iaf+ (37) & 1.9, 2.1 (37,38) & 3.41161  $\pm$ 0.00005 (39) &    &  13.8 (39)  \\
Vela X--1     & B0.5Ib & 1.8  &  8.9 (40)  &  283  (40) &  \\
4U1907$+$09     & O8-O9 Ia (41)     & 2-6    &  8.38  (42)   &  437.5 (43)    &  \\
\hline
\end{tabular}
\label{tab:sources}

(1) \citealt{Negueruela2006b};    
(2)  \citealt{Rahoui2008b};    
(3) \citealt{Drave2010}; 
(4) \citealt{Pellizza2006}; 
(5) \citealt{Clark2009};  
(6)  \citealt{Drave2012};
(7) \citealt{Negueruela2006:aTel831} 
(8) \citealt{Torrejon2010}; 
(9) \citealt{Negueruela2008int};  
(10) \citealt{Zurita2009}; 
(11)  \citealt{Bird2009};  
(12)   \citealt{Chaty2008};  
(13)   \citealt{Corbet2006};  
(14)   \citealt{Sidoli2012};  
(15)   \citealt{Corbet2013};  
(16)   \citealt{Drave2013Atel};  
(17)   \citealt{Nespoli2008};  
(18)   \citealt{Jain2009}; 
(19)   \citealt{Romano2009}; 
(20)  \citealt{RahouiChaty2008}; 
(21)   \citealt{Levine2006}; 
(22)   \citealt{Sguera2007}; 
(23)   \citealt{Coe1996};  
(24)   \citealt{Goossens2013}; 
(25)   \citealt{Clark2010};  
(26)   \citealt{LaParola2010};  
(29)   \citealt{Leyder2007};  
(30)   \citealt{Negueruela2005hd}; 
(31)   \citealt{Lorenzo2010};  
(32)   \citealt{Masetti2006};
(34)   \citealt{SidoliPM2006};  
(35) \citealt{Sidoli2007};  
(36)  \citealt{Swank2007}; 
(37) \citealt{Ankay2001};
(38) \citealt{Megier2009};
(39) \citealt{Hong2004};
(40) \citealt{McClintock1976};
(41) \citealt{Cox2005};
(42) \citealt{Marshall1980};
(43) \citealt{Makishima1984}.

\end{table*}

The INTErnational Gamma-Ray Astrophysics Laboratory, {\it INTEGRAL} \citep{winkler03,winkler11}, is a medium sized ESA mission successfully launched in 2002. 
Thanks to its large field of view, sensitivity at hard X-rays
and observing strategy, it is optimized to surveying the hard X--ray sky. It has currently 
discovered about six hundred new sources\footnote{See e.g., http://irfu.cea.fr/Sap/IGR-Sources/}.
Among these, several display extreme transient behavior,
with recurrent bright X--ray flares (reaching peak luminosities
around 10$^{36}$-10$^{37}$~erg~s$^{-1}$) of short duration 
\citep[from a few minutes to a few hours,][]{Sguera2005, Negueruela2006a}. 
These luminous and brief X--ray flares compose major outbursts lasting a few days 
 \cite[with the brightest accretion phase typically lasting $\sim$ one day or less;][]{Romano2007},  spaced by 
a very variable time interval, ranging  from a few weeks to several months \cite[e.g.][]{Blay2008}.

These hard transients, optically associated with early-type supergiant 
companions,
were called Supergiant Fast X-ray Transients (SFXTs).
SFXTs are High Mass X--ray binaries (HMXBs) hosting a compact object,
normally a neutron star, accreting mass
from the wind of the supergiant donor which under-fills its
Roche lobe \cite[see][for a recent review]{sidoli2013}.
The most controversial issues related to SFXTs deal with two aspects:  the physical 
mechanism producing
their sporadic transient X--ray emission during outbursts (which remain 
unpredictable, except
for the periodic SFXT IGR~J11215--5952, \citealt{SidoliPM2006});
 the link with more classical HMXBs, discovered in the early days
of X--ray astronomy, where the X--ray emission is persistent 
(e.g. Vela X--1, 4U~1700--377).

In orbit since 2002, \inte~allowed us to build a large
database \citep{paizis2013} which enables a statistical approach to explore
the SFXT extreme phenomena and their link to the more classical HMXBs.

In this paper, we used the long based \inte~archival data of all known 
SFXTs to fully characterize
for the first time their hard  X--ray (17--100 keV) transient emission, by means of the
cumulative luminosity distribution of their SFXT flares. 
We compared their luminosity distribution
with those of three classical HMXB systems, two persistent and one 
transient. The sources studied in this work and their main properties can be seen in Table~\ref{tab:sources}.

For clarity, we classified the sources considered in the present study into various sub-types
(\lq\lq prototypical\rq\rq, \lq\lq intermediate\rq\rq, \lq\lq less explored\rq\rq, and \lq\lq periodic\rq\rq), depending on the source
properties found in the literature.
With \lq\lq prototypical SFXTs\rq\rq~we mean the SFXTs which have displayed in the past a very high dynamic range
(ratio between the flares luminosity at their peak and the quiescent luminosity), exceeding
10$^{3}$; with \lq\lq intermediate SFXTs\rq\rq~we mean the SFXTs with a much lower dynamic range of about two orders
of magnitude (e.g. \citealt{Sguera2008}). We designate \lq\lq less explored SFXTs\rq\rq~the sources poorly
studied in the previous literature, while with \lq\lq periodic\rq\rq~we mean the only SFXT with a strictly
periodic flaring behaviour, IGR~J11215--5952 \citep{SidoliPM2006}.

\begin{table*}
 \centering
  \caption{Global view of our \ima~analysis results (exposure times are rounded for clarity). No value (\lq\lq -\rq\rq) means that the source was never detected in the given energy band.}
  \begin{tabular}{@{}ccccccc@{}}
\hline
   Name & Source $\leq12\degmark$ & Average rate$^{(1)}$ & Detections$^{(1)}$ & Detections$^{(1)}$ & Detections$^{(1)}$ & Detections$^{(1)}$\\
	&  & (17--30\,keV) & (17--30\,keV)      &  (30--50\,keV)           &  (17--50\,keV) &  (50--100\,keV) \\     
\hline
        &     Ms  (\#ScWs)  & \countsec     &   ks (\#ScWs)         &    ks (\#ScWs) &  ks (\#ScWs)&  ks (\#ScWs)\\
\hline
Prototypical SFXTs &   \\
\hline
XTE~J1739--302    & 14.6 (9215) &  11.7$\pm$0.1     & 130.7 (70)  &95.8 (52)  &  157.1 (87)  &2.5 (1)\\
IGR~J17544--2619  & 14.3 (9042) &  11.2$\pm$0.2     & 103.8 (62)  & 16.1 (9)  &  102.6 (64)  & 1.1 (1)\\
SAX~J1818.6--1703 & 7.3 (4991)  &  12.4$\pm$0.2     & 52.1 (34)   &24.7 (15)  &  73.5 (46)   &   5.0 (2)\\
\hline

Intermediate   SFXTs           &          &         &      &    & \\
\hline
IGR~J16418--4532 & 8.0 (5337)   &  9.0$\pm$0.2      & 71.9 (40)    & 10.5 (9)    & 98.0 (57)     & - \\
IGR~J16479--4514  & 7.9 (5255)  &  8.9$\pm$0.1      &  188.5 (111) &  121.0 (75) &   291.2 (172) & 7.2 (4)\\
IGR~J18483--0311 &  6.0 (4281)  &  8.4$\pm$0.1      & 199.5 (126)  &  110.8 (70) & 313.5 (200)   & 1.1 (1)\\
IGR~J18450--0435 &  5.7 (3887)  &  5.7$\pm$0.3      &  13.1 (8)    & 5.2 (4)     & 24.0 (16)     &   - \\
\hline

Less explored SFXTs      &           &           &           &         &  \\
\hline
IGR~J16465--4507  &  7.8 (5185) &   5.9$\pm$0.4      & 9.8 (6)     & 3.7 (2)         &  19.8 (11)         & -  \\
IGR~J08408--4503  &  4.5 (2343) &   4.7$\pm$0.3      &7.3     (3)  & 4.9 (2)         &  4.9 (2)       &-  \\
IGR~J18410--0535  &  5.3 (3887) &   8.0$\pm$0.3      &23.2    (16) & 14.6 (12)       &  38.4 (29)      & -  \\
\hline

The periodic SFXT &             &                &             &                 &              &  \\
\hline
IGR~J11215--5952  &  4.1 (2052) &   6.9$\pm$0.3      & 17.8 (11)   & 15.0 (9)        &  35.7 (19)   & -  \\
\hline
\hline
 HMXBs            &             &                &             &             &    \\
\hline
4U~1700--377      & 11.0 (7048) &   50.91$\pm$0.03    & 8109.96 (5105)  &  7837.87 (4922) &  8348.69 (5285) &  5565.39 (3373) \\
Vela~X--1         & 4.9 (2488)  &    58.97$\pm$0.03   & 3872.72 (1961) &  3819.02 (1927) &  3886.44 (1970) &  832.3 (363) \\
4U~1907$+$09      &  6.9 (4147) &   6.19$\pm$0.03     & 1795.38 (917)  &  80.2 (35)      &  1600.32 (815)  & - \\
\hline
\hline
Crab 		  & 4.415 (2897)&  172.70$\pm$0.03    &  4415 (2897)   &  4415 (2897)    &  4415 (2897)    &  4415 (2897)\\
\hline
\end{tabular}
\label{res:ima}

$^{(1)}$ We provide the exposure time, as well as the number of ScWs (\#ScWs), in which an 
\ima~significance $>$5 is obtained.\\

\end{table*}

 	 \section{Data Analysis: an \textit{INTEGRAL} archive \label{golia}}

The {\it INTEGRAL} payload consists of two main gamma-ray instruments, the spectrometer SPI \citep{vedrenne03} and the imager IBIS \citep{ubertini03}, covering the 15\,keV -- 10\,MeV band.  IBIS is a high angular resolution gamma-ray imager optimized  for accurate point source imaging and for the continuum and broad line 
spectroscopy. It consists of two layers,  the lower energy one 
\cite[IBIS/ISGRI, 15\,keV -- 1\,MeV,][]{lebrun03} and the higher energy one \cite[IBIS/PICsIT, 0.175--10\,MeV,][]{labanti03}. 
Co-aligned with SPI and IBIS are two X--ray monitors JEM--X \cite[4--35\,keV,][]{lund03} and an optical monitor OMC \cite[500--600\,nm,][]{mashesse03}. At the time of writing a total of about ten years of data have become public and are available to the scientific community. In order to increment and ease our exploitation of {\it INTEGRAL} data, we undertook the task of preparing and maintaining an {\it INTEGRAL}~archive \citep{paizis2013}.  The scripts we have used to build 
it are publicly available\footnote{http://www.iasf-milano.inaf.it/$\sim$ada/GOLIA.html}.

The data used in this study span from revolution 0026 to 1159, i.e., 
December 2002 -- April 2012 (90491 pointings). A detailed description of the data analysis  
and products is given in \cite{paizis2013}. 
Here  only the main information is given, for completeness.\\
\inte~data are downloaded from the ISDC Data Centre for Astrophysics and a customized 
analysis using the OSA 9.0 software package \footnote{http://www.isdc.unige.ch/integral/analysis} is routinely performed on the IBIS/ISGRI data. 

IBIS is a coded aperture imaging system. In such systems, the source radiation is spatially
modulated by a mask of opaque and transparent elements
before being recorded by a position sensitive detector. This enables
simultaneous measurement of source plus background (through the mask holes) and background
fluxes (through the opaque elements). To optimize sky image reconstruction, mask
patterns are designed so that  each source in the field of view
results in a unique shadowgram on the detector. The image 
reconstruction (deconvolution) is  based on a correlation
procedure between the recorded image and a decoding array derived 
from the mask pattern  \citep{goldwurm01,goldwurm03}. The scientific 
products include individual pointing images (pointing duration $\sim$\,ks) and the associated detected source lists in 
the selected energy bands. Hereafter we refer to the image deconvolution results as to \ima~results.\\
Once the positions of the active sources of the field are known from the \ima~step,
their fluxes are extracted for each pointing in predefined energy and time bins. This extraction is based on simultaneous fitting of source and background shadowgram models to detector images. Hereafter we refer to the lightcurve extraction  results as to \lcr~results.

The scientific products obtained in our analysis include individual pointing images and the associated detected source lists in the 17--30, 30--50, 17--50 and 50--100\,keV energy bands (\ima~results), as well as light-curves binned over 100\,s in the 17--30\,keV band for sources detected in the images (\lcr~results).  

We consider a 5$\sigma$ detection threshold in the \ima~step (in each band), and a  3$\sigma$ 
detection threshold in the \lcr~one (where we know the source to be active from \ima). For SFXTs alone, as a cross-check, we have extracted \ima~results in the 22--50\,keV band as well, to make sure that the instrumental low threshold fluctuations do not introduce any kind of bias in our results.   \\

In this work we show the overall behaviour of the selected sources in the hard X--ray domain spanning over about  nine years, mapping the behaviour of the sources in two time-frames: pointing basis ($\sim$\,ks, \ima~results) and light-curve  basis (100\,s, \lcr~results). We have considered only the pointings (or Science Windows, hereafter ScW) in which the sources were within $12\degmark$~from the centre. Indeed, at larger off axis angles the IBIS response is not well known and strongly energy dependent, hence systematic flux variations may be introduced\footnote{http://www.isdc.unige.ch/integral/osa/9.0/issues\_osa}. A quick glance at the data between 12--20$\degmark$ is however included in section~\ref{result}, for completeness.

Besides the sources shown in Table~\ref{tab:sources}, we include  the results from the Crab source, that we consider like the \lq\lq point spread function\rq\rq~of our archive, to check for the consistency of our analysis. \\

  	\section{Detections and duty cycles}\label{result}

\subsection*{\ima~results ($\sim$ks sampling)}\label{sec:ima}

In Table~\ref{res:ima} we present our results from the \ima~step. The total exposure time for which each source was within $12\degmark$~is given, together with the exposure time (and number of ScWs) in which the source was found to be active in each band (\ima~significance $>$5). Most sources have the highest detection rate in the 17--50\,keV band, hence  this band is the best one to investigate the duty cycles of SFXTs in hard X--rays.  Table~\ref{res:duty} shows the computed duty cycles for the sources in our sample. We have included the 17--30\,keV band as well, for comparison.

\begin{table}
 \centering
  \caption{Obtained \ima~duty cycles in the 17--50 and 17--30\,keV band.}
  \begin{tabular}{ccc}
\hline
   Name & Duty cycle (\%)$^{(1)}$ &  Duty cycle (\%)$^{(1)}$  \\
	&  (17--50\,keV) & (17--30\,keV)  \\ 
\hline
Prototypical SFXTs & & \\
\hline
XTE~J1739--302  	& 1.07		& 0.89 \\
IGR~J17544--2619  	& 0.72 		& 0.72 \\
SAX~J1818.6--1703 	&1.00 		& 0.71\\
\hline
Intermediate  SFXTs           & &  \\
\hline
IGR~J16418--4532 	& 1.23 		&  0.90\\
IGR~J16479--4514 	&3.68 		& 2.39\\
IGR~J18483--0311 	& 5.22 		& 3.31 \\
IGR~J18450--0435 	& 0.42		& 0.23 \\
\hline
Less explored SFXTs	& \\
\hline
IGR~J16465--4507 	&0.25 		& 0.13 \\
IGR~J08408--4503  	& 0.11 		& 0.16 \\
IGR~J18410--0535        &0.72		& 0.44 \\
\hline
The periodic SFXT      &   &  \\
\hline
IGR~J11215--5952  	&0.87		& 0.44 \\
\hline
\hline
HMXBs           & &  \\
\hline
4U~1700--377 		& 76.0 		& 73.8 \\
Vela~X--1     		& 79.4 		& 79.1\\
4U~1907$+$09     	&23.1 		& 26.0\\
\hline
\hline
Crab 			&  100 		& 100\\
\hline
\end{tabular}
\label{res:duty}

$^1$ Duty cycle (i.e. percentage of source activity computed  using the exposure times of Table~\ref{res:ima}).
\end{table}

\subsection*{\lcr~results (100\,s sampling)} \label{sec:lcr}
\begin{table}
\centering
\caption{Global view of our \lcr~analysis results.}
\begin{tabular}{cccc}
\hline
   Name & Source $\leq12\degmark$ $^{(1)}$ & Detections$^{(2)}$& Duty cycle$^{(3)}$   \\
	&  & (17--30\,keV) &  (17--30\,keV)  \\     
 \hline
        &       \#bins  &   \#bins  &   \%   \\
\hline
Prototypical SFXTs &          \\
\hline
XTE~J1739--302  & 1938 & 530 & 27.3 \\
IGR~J17544--2619  & 1589  & 334 & 21.0\\
SAX~J1818.6--1703 & 769  & 207 & 26.9 \\
\hline
Intermediate SFXTs           &    \\
\hline
IGR~J16418--4532 &  1097  & 77 & 7.0 \\
IGR~J16479--4514  & 2856  &319 & 11.2 \\
IGR~J18483--0311 & 3034   & 397 & 13.1\\
IGR~J18450--0435 &  200  & 8  &  4.0\\
\hline
Less explored SFXTs     \\
\hline
IGR~J16465--4507  &  142   & 7  & 4.9\\
IGR~J08408--4503  &  103   & 17 & 16.5 \\
IGR~J18410--0535  &  348   & 60 & 17.2\\
\hline
The periodic SFXT   \\
\hline
IGR~J11215--5952 & 254  & 40 & 15.7 \\
\hline
\hline
 HMXBs           &        \\
\hline
4U~1700--377 & 123337  & 84693 & 68.7\\
Vela~X--1     & 57543 &  52679 & 91.5 \\
4U~1907$+$09     & 26154 & 3508 & 13.4 \\
\hline
\hline
Crab &   68543  & 68246 & 99.6 \\
\hline
\end{tabular}
\label{tab:lcr_3s}

$^{(1)}$ Not all bins have exactly 100\,s (the last bin of each ScW will be less), hence the correct exposure times of the source within 12$\degmark$ are to be taken from Table~\ref{res:ima}.\\
$^{(2)}$ Bin with $\sigma >$ 3.\\
$^{(3)}$ Duty cycle computed using the number of bins.\\

\end{table}

 Once a source is detected in the imaging part, i.e. active in a ScW, a lightcurve with a 100\,s binning is also extracted in the 17--30\,keV band.
In Table~\ref{tab:lcr_3s} we present our results from the light-curve step. The  number of 100\,s bins in which the source was within 12$\degmark$ (and active in \ima) is given, together with the number of bins in which there is at least a 3\,$\sigma$ detection in a single 100\,s bin. The duty cycles (i.e. percentage of source activity at the 100\,s level) are also shown. We note here that the results from the light-curve step show the short term variability \textit{when the source is active at the ScW level}, hence the duty cycles given are not with respect to the whole nine year archive, but with respect to when the source was active in the \ima~step (e.g. XTE~J1739--302 is detected 27.3\% of the times in 100\,s bins during its \ima~activity). 

\subsection*{The outer part of the field of view}\label{beyond}
Since spectral analysis and conversion to luminosities are involved in our work, we decided to present our results only within $12\degmark$, where the IBIS/ISGRI response is better known and where the signal to noise reconstruction is cleaner. A detection of a source beyond $12\degmark$, i.e. in the most extreme part of the detector, could be fake and in an archival approach such as ours, it could be misleading.  Nevertheless, for completeness, we briefly report here the detections in the outer part of the field of view, namely between $12\degmark$ and $20\degmark$. 

The number of ScWs in which the sources are detected at least at 5$\sigma$ in the 17--50\,keV band between $12\degmark$ and $20\degmark$ in the images are as follows: 
14 for IGR~J18483--0311, 
12 for \sax, 
5 for IGR~J16479--4514, 
2 for XTE~J1739--302, 
2 for IGR~J17544--2619,  
1 for IGR~J16418--4532,
1 for IGR~J11215--5952,
1 for IGR~J18450--0435,
 and 
0 for IGR\,J16465--4507,  IGR~J08408--4503 and IGR~J18410--0535. A detailed analysis of these pointings to assess their true nature is beyond the scope of this paper. \\
The same $12\degmark$ radius choice is applicable to the \lcr~part. Indeed, once we consider a 20$\degmark$ radius, we obtain luminosity distributions unrealistically stretched towards very high luminosities: the Crab distribution  is detected at more than 3$\sigma$ up to 1$\times$10$^{37}$ \ergsec.

	      \section{From count-rates to luminosity} \label{sec:spectra}
Conversion factors from IBIS/ISGRI count-rates to X--ray luminosities (in the same
energy range) have been derived from the analysis of IBIS/ISGRI spectra. 
We have extracted spectra from seven ScWs (or less for the two sources where seven 
detections were not reached) in which each source was within $12\degmark$~from the centre and detected above 5\,sigma in the \ima~step.
We verified that the spectra of these sources did not show strong evidence for variability
and fitted the average spectra to obtain reliable conversion factors from source
count-rates to X--ray fluxes, in the same energy range.
Different kinds of models have been adopted, in order to get a good deconvolution of the
average spectra.
We assumed the appropriate source distances to obtain the relative luminosities in each  energy range.
The final conversion factors for each band and the distances used are given in Table~\ref{tab:conv}. 
We note that the main uncertainty in this process is relative to the source distances.

	      \section{Cumulative luminosity distributions}\label{sec:distributions}
Merging the results shown in the two previous sections, for each source we built the \textit{complementary cumulative distribution function} (hereafter, only
{\em cumulative distribution}) of the obtained X--ray luminosities. In each point of these functions,
at a given luminosity L$_{X}$ the sum of all events (detections) with a
luminosity larger than L$_{X}$ are plotted.
The advantage of the cumulative distributions is that there is no
need to arbitrarily bin the data, enabling a comparison between the different sources
and avoiding loss of information.

\begin{figure*}
\begin{center}
\centerline{\includegraphics[width=14cm]{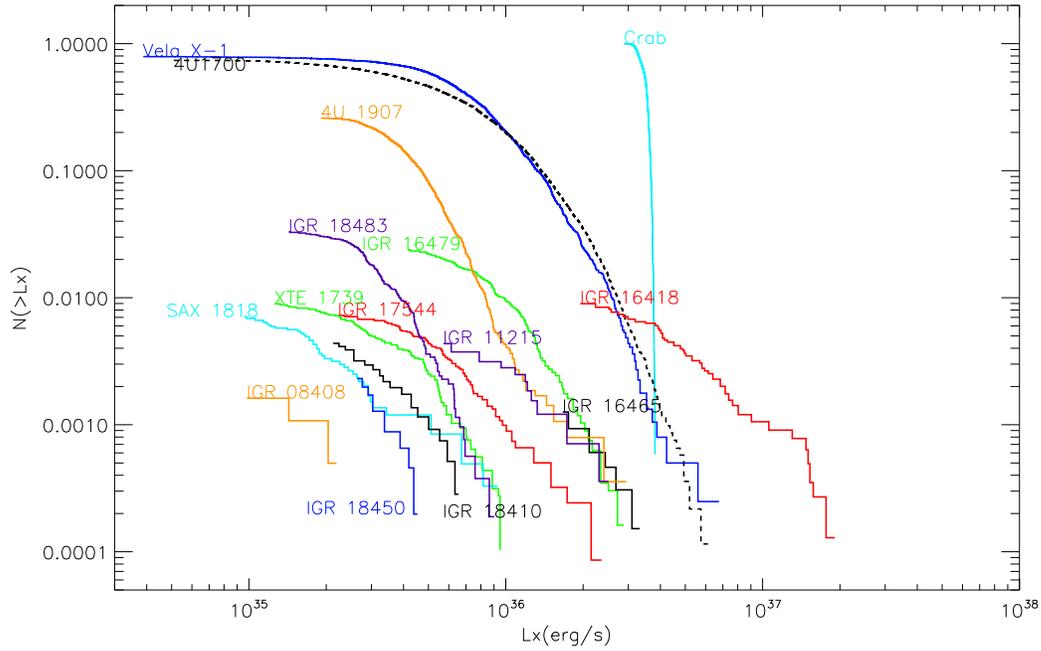}} 
\caption{Cumulative luminosity distributions for all the sources of our sample in 17--30\,keV, as obtained from the imaging analysis ($\sim$ks bins). Each curve has been normalized to the total exposure time for which the source was within $12\degmark$ (Table~\ref{res:ima}). The duty cycles, the highest values in the y-axis, are with respect to the whole archive and are given in Table~\ref{res:duty} for the 17--30\,keV band  shown here and for the 17--50\,keV band shown in Figure~\ref{fig:ima3}.}
\label{fig:ima1}
\end{center}
\end{figure*}

\begin{figure*}
\begin{center}
\centerline{\includegraphics[width=14cm]{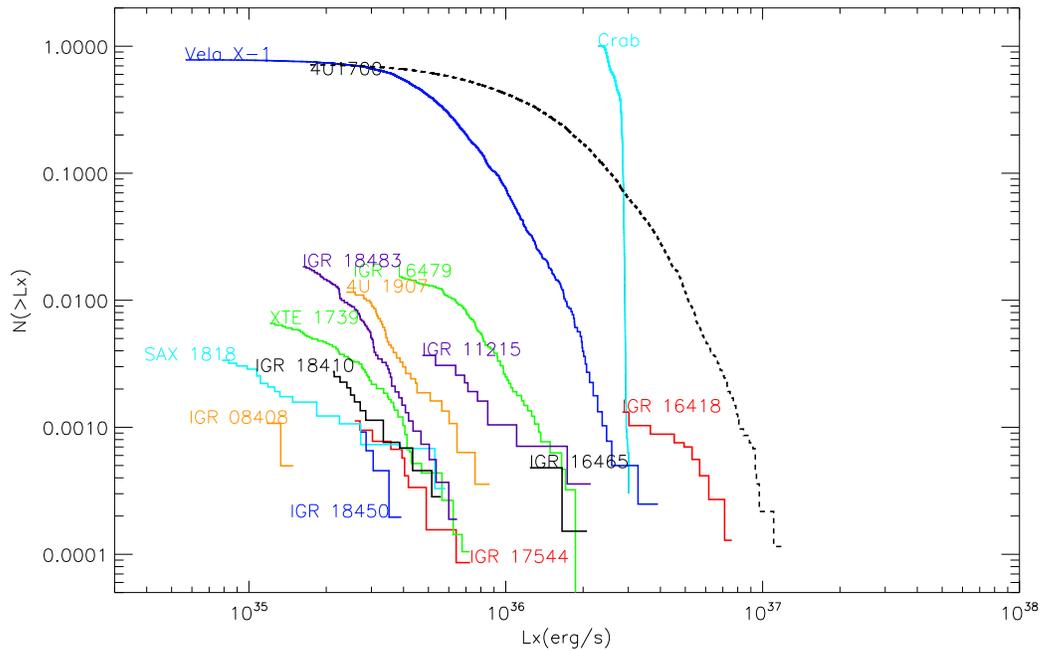}} 
\caption{As in Figure~\ref{fig:ima1} but for 30--50\,keV. }
\label{fig:ima2}
\end{center}
\end{figure*}

\begin{figure*}
\begin{center}
\centerline{\includegraphics[width=14cm]{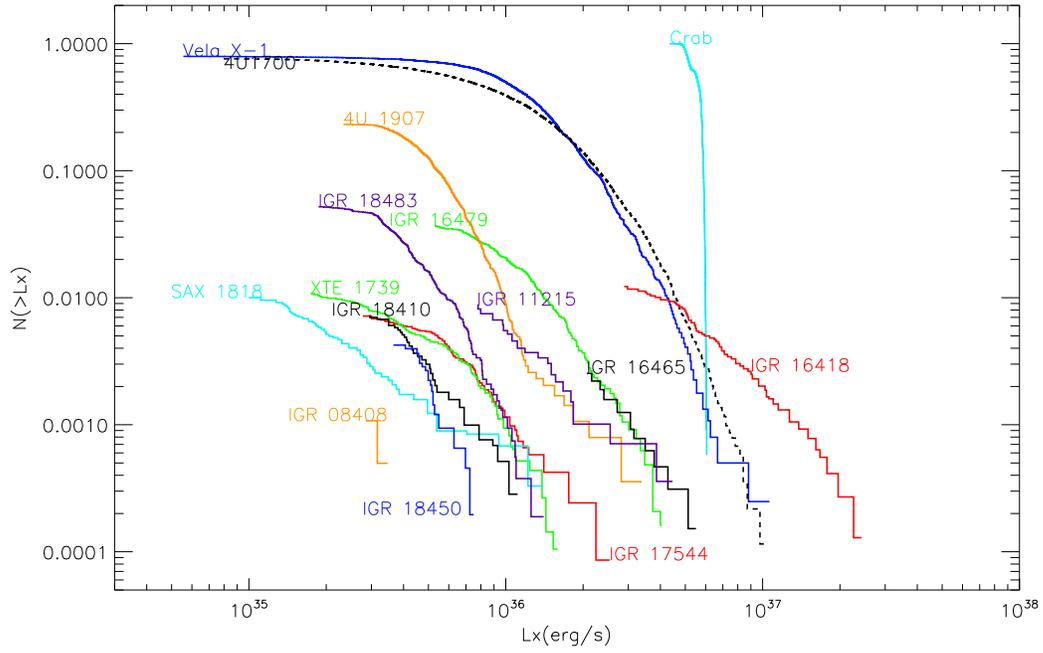}} 
\caption{As in Figure~\ref{fig:ima1} but for  17--50\,keV. }
\label{fig:ima3}
\end{center}
\end{figure*}

\begin{figure*}
\begin{center}
\centerline{\includegraphics[width=14cm]{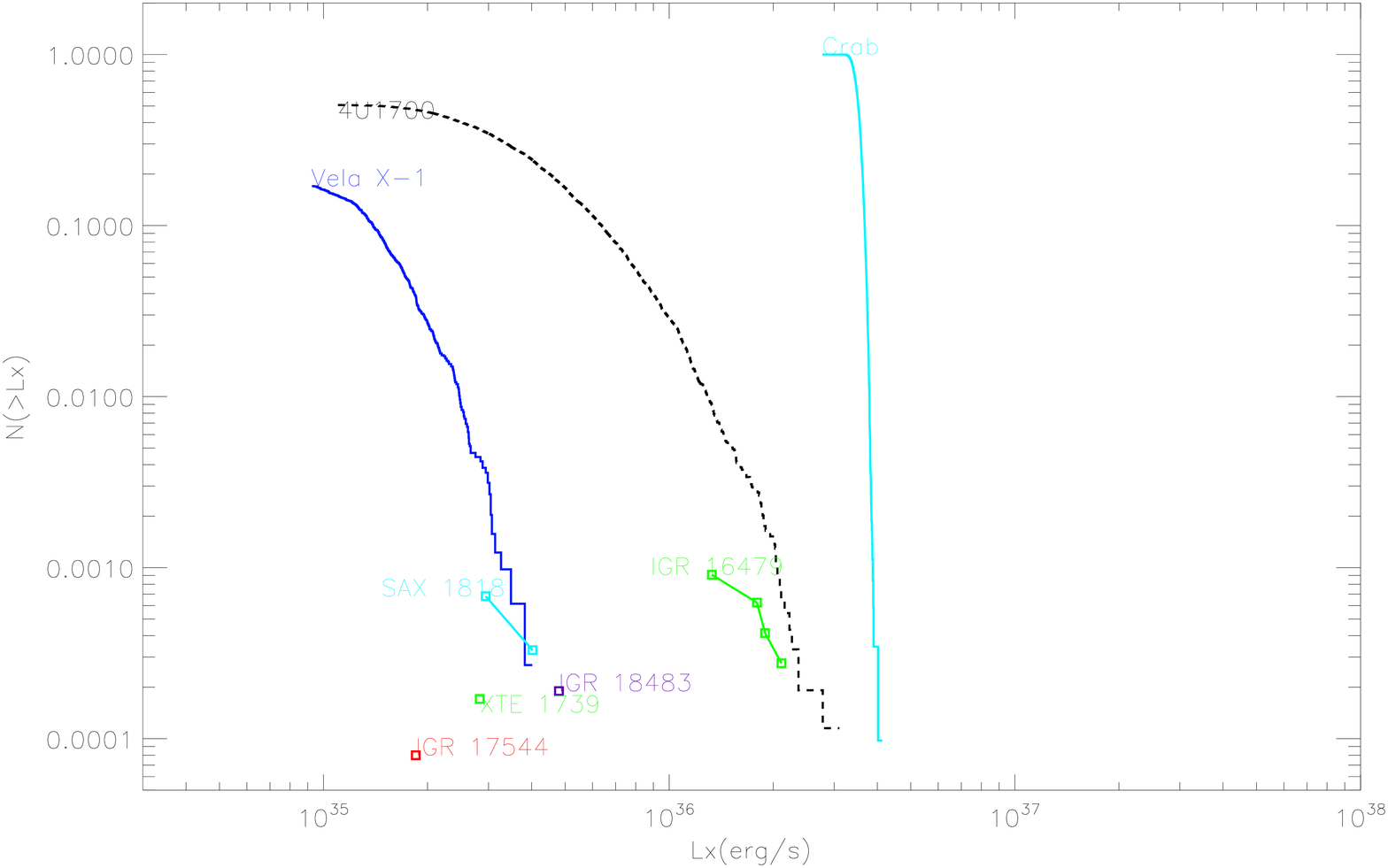}} 
\caption{As in Figure~\ref{fig:ima1} but for  50--100\,keV.}
\label{fig:ima4}
\end{center}
\end{figure*}

Figures~\ref{fig:ima1} to ~\ref{fig:ima4} show the resulting cumulative distributions for all the sources of
our sample, in all the extracted bands. Each curve has been normalized to the total exposure time for which the source was within $12\degmark$. The source duty cycles, i.e. the percentage of time where the source is active with respect to the whole database,
can be seen in each plot as the highest value in the y-axis, and is given in Table~\ref{res:duty} for the 17--50 and 17--30\,keV bands.  As can be seen, SFXTs have a few percent duty cycle and are very soft,  basically disappearing in the 50--100\,keV band (Figure~\ref{fig:ima4}).

In these plots, a perfectly constant source detected by an ideal detector would result in a vertical line. In our case, we see that the Crab source has a deviation from the straight line, towards lower luminosities. This is due to the combination of two effects: instrumental and intrinsic to the source.  In a nine year operational period, IBIS/ISGRI response has  changed, hence the resulting Crab lightcurve is not constant. The overall Crab count rate trend in 17--30\,keV as obtained from the \ima~analysis can be seen in Figure~\ref{fig:lcr1_3s}, right panel, in black. This effect, however, is clearly visible in the Crab because it is very bright, while it is washed away by the statistics in the other sources.
Our overall observed variability (i.e. deviation from a vertical straight line in Figures 1 to 4) also includes the intrinsic $\sim$7\% flux decline that has been observed in the hard X--ray emission of the Crab, independently confirmed by several instruments, including IBIS in the 15--50\,keV band \citep{wilson11}. 
In the cumulative distributions no time information is retained, hence we cannot disentangle the two   variabilities (instrumental versus intrinsic).

\subsection{Comparing slow and fast variability}\label{sec:compare}
\begin{figure*}
\begin{center}
\centerline{\includegraphics[width=1.0\linewidth]{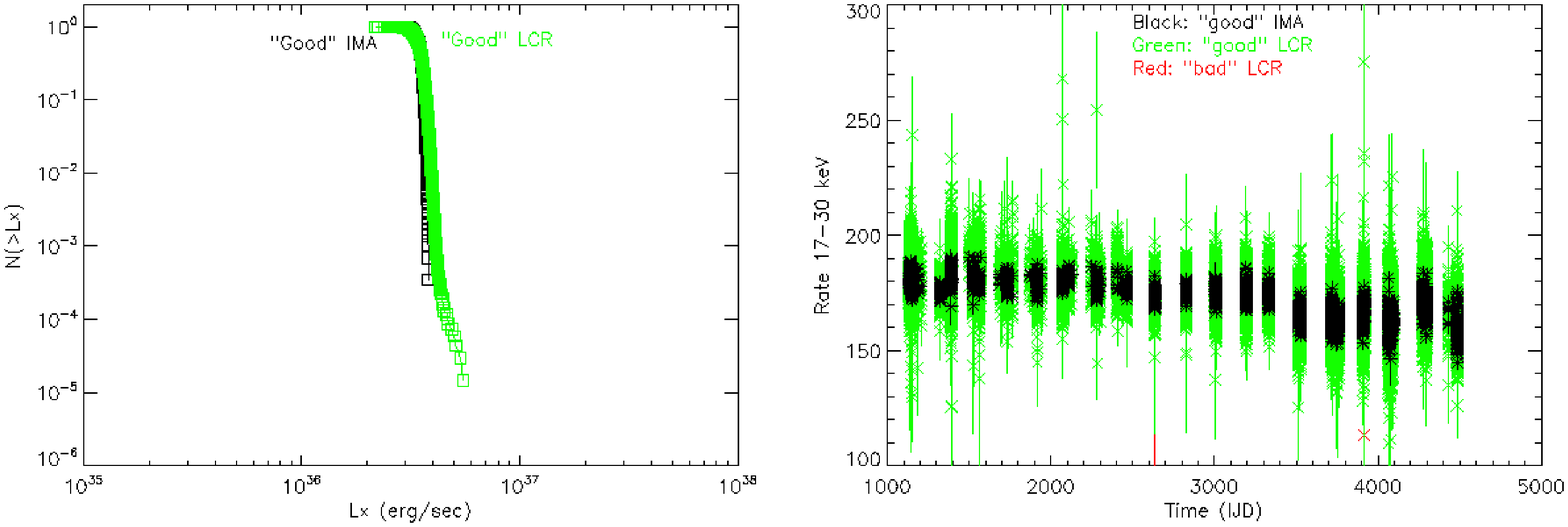}} 
\caption{\emph{Left panel}: comparison of the Crab luminosity cumulative distributions, as obtained from \ima~step (black) and \lcr~(green). A flatter slope indicates more variability.  \emph{Right panel}: Crab lightcurve in the 17--30\,keV band. Black: \lq \lq good \rq \rq~rate and error bars from \ima~step (i.e. $\sim$ks bin detection above 5$\sigma$ and off axis angle $<12\degmark$). Green:  \lq\lq good\rq\rq~rate and error bars from the \lcr~step (i.e. 100\,s bin detection above 3$\sigma$ within the \lq\lq good\rq\rq~\ima~ScWs). Red: \lq\lq bad\rq\rq~rate and error bars from \lcr~step (i.e. detection below 3$\sigma$). }  
\label{fig:lcr1_3s}
\end{center}
\end{figure*}

\begin{figure*}
\begin{center}
\centerline{\includegraphics[width=1.0\linewidth]{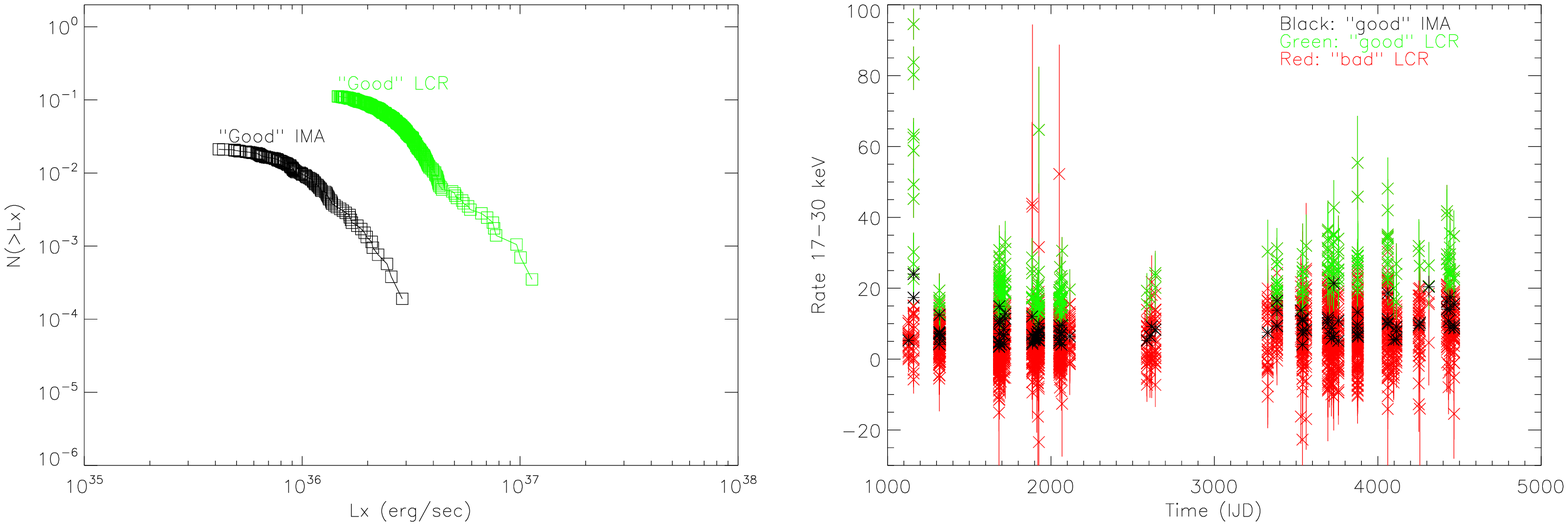}} 
\caption{As in Fig~\ref{fig:lcr1_3s} but for IGR~J16479--4514.}
\label{fig:lcr2_3s}
\end{center}
\end{figure*}

\begin{figure*}
\begin{center}
\centerline{\includegraphics[width=1.0\linewidth]{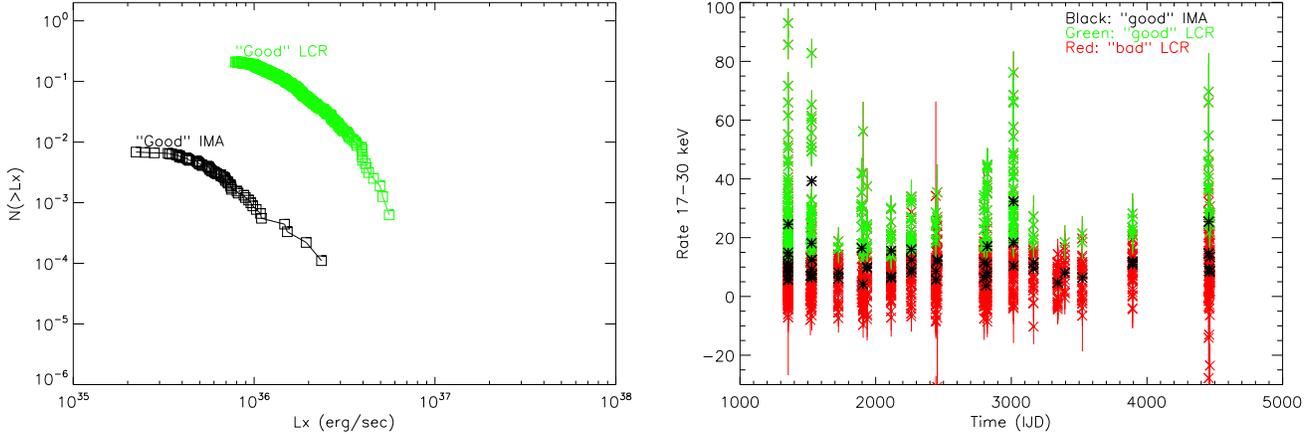}}
\caption{As in Fig~\ref{fig:lcr1_3s} but for IGR~J17544--2619.}
\label{fig:lcr3_3s}
\end{center}
\end{figure*}

To have a better feeling of how the obtained distributions change according to the binning chosen ($\sim$\,ks versus 100\,s in 17--30\,keV), we show some examples where we plot in the same diagram the two different curves belonging to the same source. Figures~\ref{fig:lcr1_3s},~\ref{fig:lcr2_3s} and~\ref{fig:lcr3_3s}, left panels, show the Crab, IGR~J16479--4514 and IGR~J17544--2619, respectively. As it can be seen, in the case of the Crab the two curves (black for ScW \ima, and green for 100\,s \lcr) basically overlap. This is because the Crab is bright and stable, hence if we  look at its lightcurve (Figure~\ref{fig:lcr1_3s}, right panel), we  see that the distribution of the 100\,s rates (green) is about symmetrically placed around the obtained, average, ScW value (black). Furthermore, the finer binning has a larger statistical scatter and this results in a broader coverage of the corresponding distribution, visible in the left panel.
We note that in the \lcr~based distribution, unlike the \ima~based one, the Crab has a small high energy tail that deviates towards higher luminosities. This is due to six 100\,s bins (out of a total of 68246) that reach about 250\,\countsec~(Figure~\ref{fig:lcr1_3s}, right panel, in green). We expect the Crab to be the worst case scenario and even here the percentage of bins that deviate is negligible.

For IGR~J16479--4514 and IGR~J17544--2619, the two cumulative distributions do not overlap (Figures~\ref{fig:lcr2_3s} and~\ref{fig:lcr3_3s}, left panels). Indeed the 100\,s cumulative curves (green in the left panels) lie at higher luminosities than those obtained from the ScW average (black). This is because the sources are fainter and lie closer to the detection threshold, hence, the bins in which we consider a detection, $>3\sigma$, are  the brightest in the corresponding light-curves (green in the right panels). On the contrary, the bins with detection lower than $3\sigma$ are rejected (shown in red) and are not included in the \lcr~curves of the left panels, but they still contribute to build the average \ima~value (black). While the shift of the cumulative distributions to higher luminosities (increasing x-axis) from \ima~to \lcr~is due to the $3\sigma$ selection effect, the shift towards a higher percentage (increasing y-axis) is due to the fact that the \lcr~duty cycles are not with respect to the whole nine year archive, as \ima, but with respect to when the source was active in the \ima~step (i.e. \lcr~duty cycles are higher than the \ima~ones).

 \begin{table*}
 \centering
  \caption{Conversion factors from count-rates (\countsec) to luminosities (\ergsec). No value (\lq\lq -\rq\rq) means that the source was never detected in the given energy band (namely 50--100\,keV, see Table~\ref{res:ima}); \lq\lq n/a\rq\rq~means that the source was not extracted in the given band (only SFXTs have been extracted in 22--50\,keV, see section~\ref{sec:newrun}). }
 \begin{tabular}{ccccccc}
\hline
   Name & Distance  & 17--30\,keV & 30--50\,keV & 17--50\,keV & 22--50\,keV  & 50--100\,keV\\
	& (kpc)	    & ($\times$ 10$^{34}$)& ($\times$ $10^{34}$)& ($\times$ $10^{34}$)&($\times$ $10^{34}$)& ($\times$ $10^{34}$)\\	
\hline

Prototypical SFXTs &          &          &           &    & \\
\hline
XTE~J1739--302 &  2.7    & 3.50 & 5.20 & 4.00 & 3.10 & 9.60  \\
IGR~J17544--2619 & 3.6   & 6.00 & 10.5 & 6.30 & 6.20 & 4.05 \\
SAX~J1818.6--1703 & 2    & 2.00 & 2.80 & 2.20 & 2.00 & 4.10 \\%
\hline

Intermediate  SFXTs &          &          &         &      &   \\
\hline
IGR~J16418--4532 & 13    & 75.0 & 100.0 & 90.0 & 86.0& -\\
IGR~J16479--4514 & 4.9   & 12.0 & 17.0  & 13.0 & 14.0& 35.0 \\
IGR~J18483--0311 & 3     & 4.30 & 6.50 & 5.00 & 5.40 & 11.0 \\
IGR~J18450--0435 & 3.6   & 6.20 & 10.0 & 7.40& 7.80& - \\
\hline

Less explored SFXTs      &           &           &           &         &  \\
\hline
IGR~J16465--4507 &   9.5 & 41.0 & 54.0 & 49.0 & 46.0 & -  \\
IGR~J08408--4503  & 2.7  & 3.57 & 4.98 & 3.94 & 4.14 & - \\
IGR~J18410--0535 &   3.2 & 4.95 & 7.43 & 5.66 & 5.94 & - \\
\hline

The periodic SFXT      &           &           &           &         &  \\
\hline
IGR~J11215--5952  & 6.2 & 18.6 & 26.8 & 21.0 & 22.6 & -\\
\hline
\hline 
 HMXBs  &          &          &         &     &    \\
\hline
4U~1700--377 &   1.9 &  1.60 & 6.90 & 2.00 & n/a & 4.00 \\
Vela X--1    &  1.8  &  1.40 & 2.90 & 1.90 & n/a & 4.30 \\
4U1907$+$09     & 4    &  7.60 & 15.0 & 8.70 & n/a & - \\
\hline
\hline
Crab 	& 2          &  2.00 & 3.00 & 2.20 & n/a  & 5.30\\
\hline
\end{tabular}
\label{tab:conv}
\end{table*}

	      \section{SFXTs, a new extraction in 22--50\,keV} \label{sec:newrun}

\begin{table*}
 \centering
  \caption{Results from our 22--50\,keV \ima~analysis on SFXTs. }
  \begin{tabular}{ccccccc}
\hline
Name & Average rate$^{(1)}$ &  Detections$^{(1)}$  & Minimum L${_X}$ & Maximum L${_X}$ & Mean L${_X}$& Median L${_X}$ \\
     &   (\countsec) &ks (\#ScWs)  &($\times$10$^{35}$ \ergsec) & ($\times$10$^{35}$ \ergsec) & 	($\times$10$^{35}$ \ergsec) & ($\times$10$^{35}$ \ergsec)\\

\hline
Prototypical SFXTs &          \\
\hline
XTE~J1739--302    & 10.3 $\pm$ 0.1	&135.3 (74)	& 1.1  &8.6  & 3.2 & 2.8 \\
IGR~J17544--2619  & 7.9$\pm$0.2		& 80.7 (48)	&1.7  & 15.1 & 4.9 & 4.3 \\
SAX~J1818.6--1703 & 10.1$\pm$0.2		& 55.4 (35)	&0.8 & 8.6 & 2.0 &  1.5\\
\hline
Intermediate SFXTs           &    \\
\hline
IGR~J16418--4532 & 7.2$\pm$0.2 	& 59.7 (38)	&25.4  & 159.1 & 61.9 & 54.0 \\
IGR~J16479--4514  &7.6$\pm$0.1 	& 240.4 (144) 	&4.0 &31.2 & 10.6 & 9.5\\
IGR~J18483--0311 & 7.2$\pm$0.1 	& 243.7 (156)	&1.6  & 10.2 & 3.9 & 3.5 \\
IGR~J18450--0435 & 5.2$\pm$0.2 	& 19.8 (13) 	&2.8 & 6.2 &  4.1 & 3.7\\
\hline
Less explored SFXTs      \\
\hline
IGR~J16465--4507  &4.6$\pm$0.3 	&12.4 (6) 	&14.32   & 35.2 & 21.2 & 22.8\\
IGR~J08408--4503  &5.5$\pm$0.4 	& 4.9 (2)	&2.0  &  2.5 & 2.3 & 2.5\\
IGR~J18410--0535  &6.9$\pm$ 0.2	&  32.2 (24)	&2.3  & 8.2  & 4.1& 3.6 \\
\hline
The periodic SFXT \\
\hline
IGR~J11215--5952 &5.5$\pm$0.2 	& 27.4 (15)	&5.9 &34.6 & 12.4 & 10.9 \\
\hline
\end{tabular}
\label{tab:22-50}

$^{(1)}$ We provide the exposure time, as well as the number of ScWs (\#scws), in which an 
\ima~significance $>$5 is obtained.\\

\end{table*}

We initially adopted a low energy boundary of 17\,keV to maximize the detections with IBIS/ISGRI. However, the efficiency of IBIS/ISGRI fluctuates with the lower energy threshold, so that the detector response is not stable.  
In the case of the SFXTs of our sample, about 80\% of the detections occur before revolution 848, where the low energy threshold (LT) of IBIS/ISGRI fluctuates between about 16 and 19\,keV (Caballero et al., 2012) hence for our purposes, the IBIS response can be considered basically stable. For the remaining 20\% of our detections, the efficiency of IBIS/ISGRI is not optimal in our chosen energy range and we may be underestimating the detected fluxes. We do not expect this to impact in an important way the results shown up to now: \ima~and \lcr~results have been compared in the same band, the majority of our detections are before the increase of the LT, and we use the wide and most efficient 17--50\,keV energy band to compute the duty cycles.  In any case, the effect of the LT increase in IBIS/ISGRI is  to eventually miss some detections in the 20\% portion of the data where the LT is between 20 and 22\,keV. 

We note, however, that since we intend to discuss the cumulative luminosity distributions,  as a cross-check we have re-analyzed all the ScWs for which the SFXTs have been detected in the 17--30 or 17--50\,keV band, extracting images in the 22-50\,keV band. Our results are hence free from LT fluctuations, but many detections are lost since we are not considering the 17-22\,keV bit anymore (20\% of the detections less than the 17--50\,keV range). Since this analysis considers a subsample of the whole archive (656 ScWs versus the complete archive, currently 90491 ScWs), the duty cycles for this \ima~step are arbitrary and will not be discussed any further.\\
 A global view of our results is given in Table~\ref{tab:22-50}  while the newly extracted SFXT luminosity distributions are 
discussed Section~\ref{sec:KS}. We anticipate here that, besides the percentage of detections, there is no qualitative difference between the 17--50\,keV and 22--50\,keV results, as far as our discussion is concerned.

        \section{Characterizing the cumulative distributions }\label{sec:KS}

After building the cumulative distributions, we tried to characterize
them in a more quantitative way, modeling them with  a power-law.
Note that a striking feature is present in  the cumulative distributions, 
i.e. a low luminosity turn-over. In the case of SFXTs, this is due to the missing detections of faint flares
near the sensitivity threshold of the detector, and is often observed in cumulative
distributions in many different contexts \citep{Clauset2009}.
This implies that the sampling is not complete  {\em near} this low luminosity cutoff,
and that only data points lying above a so-called {\em truncation point} 
can be considered in the estimation of the best power-law slope.

Sometimes, at the high end of the luminosity distribution,
a turn-over is observed as well.
This can be due to a \lq\lq finite\rq\rq~size scaling caused by the fact that
the system has a finite length scale (a maximum luminosity in this
case). Alternatively, it can be due to the fact that
nine years of observations are not enough to catch the most extreme flares
and cover all the statistics of the flaring activity from a given SFXT.
This implies that the upper end of the SFXTs luminosity distributions
is affected by larger uncertainties, because a few extreme events can
significantly change the high luminosity tail.

To evaluate the power-law slope of the SFXTs cumulative distributions,
we restricted our analysis to the energy band 17--50\,keV (\ima~results),
to take advantage from the wider energy coverage and statistics (and to 22--50\,keV for comparison).
We adopted a Maximum-Likelihood Estimation (MLE) of the power-law slope from a 
subsample
of data points above a truncation point, different for each source 
\citep{Crawford1970}.
Since the resulting power-law slope is strongly dependent on the assumed
lower bound, we  obtained the best truncation point empirically, by optimizing the 
Kolmogorov-Smirnov (KS) goodness-of-fit statistics \citep{Clauset2009}: 
for each source, at first a truncation point for the lowest luminosity value in the cumulative distribution 
is assumed; then a MLE method has been applied to obtain the power-law slope and
a KS test is performed. This loop is repeated increasing the value of the truncation point.
Finally, we chose as the correct value for the truncation point the one which makes the probability distributions
of the measured data and the best-fit power-law model as similar as possible
above that particular truncation point.

The results obtained with this method are reported in Table~\ref{tab:powfit} for all the sources except the
two that have too few detection (the less explored SFXTs, IGR~J16465--4507 and IGR~J08408--4503).
We list the final truncation point adopted for each cumulative distribution, 
the number of data points used (together with the initial total number of data points)  
and the best power-law slope obtained. The corresponding KS probability is also listed. 

\begin{figure*}
\centering
\begin{tabular}{cc}
\includegraphics[height=5.8cm, angle=0]{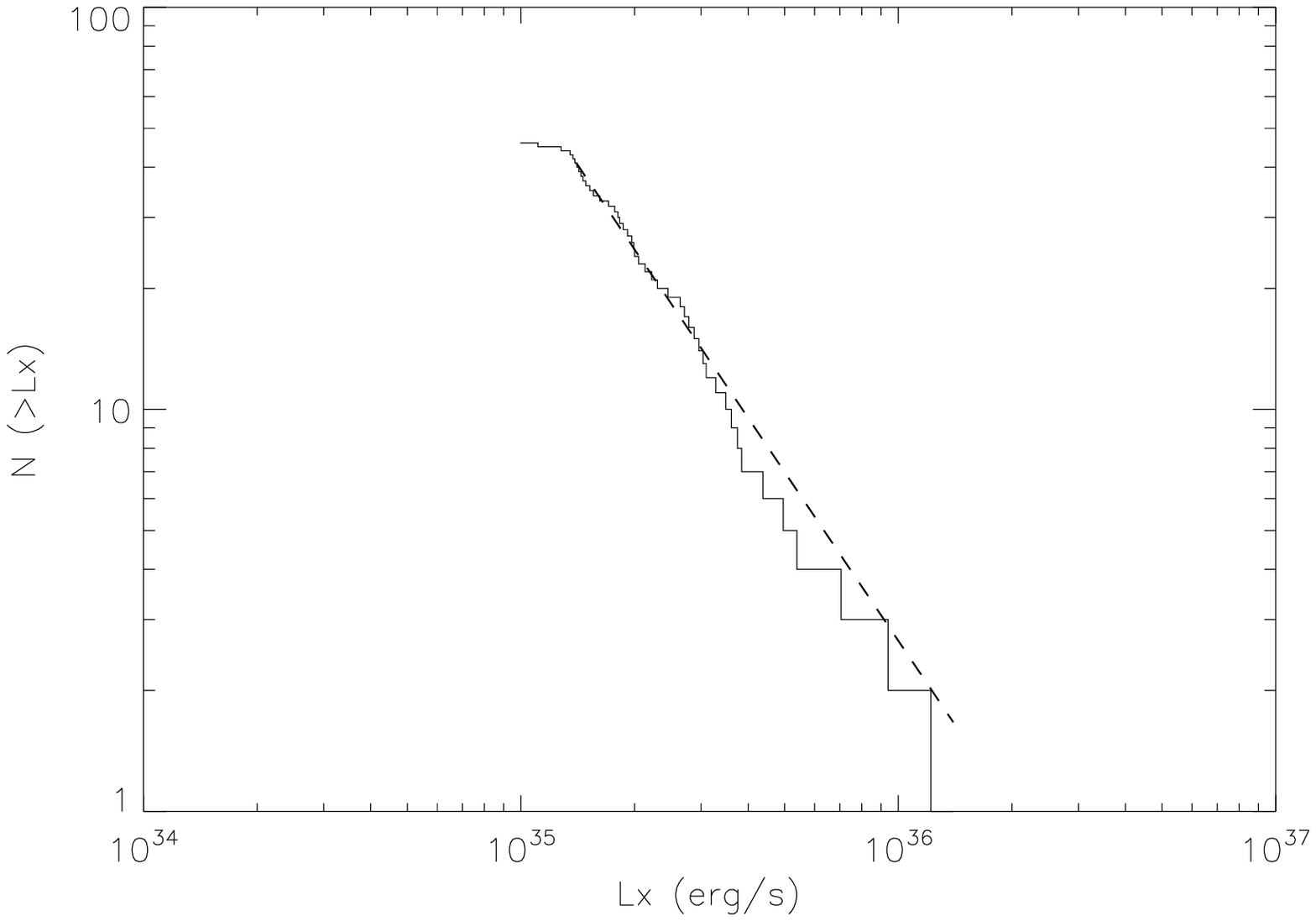} & 
\includegraphics[height=5.8cm, angle=0]{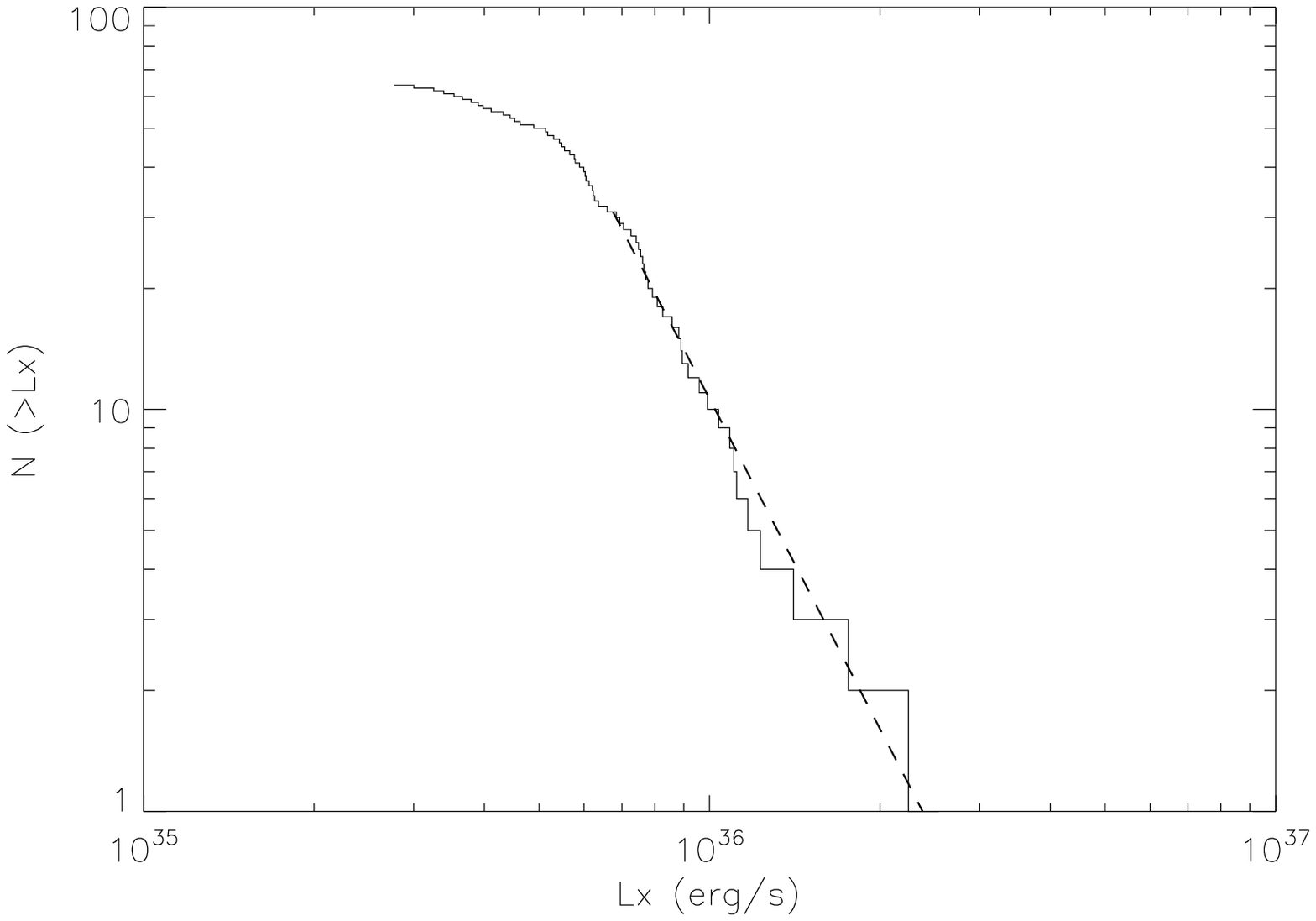} \\
\includegraphics[height=5.8cm, angle=0]{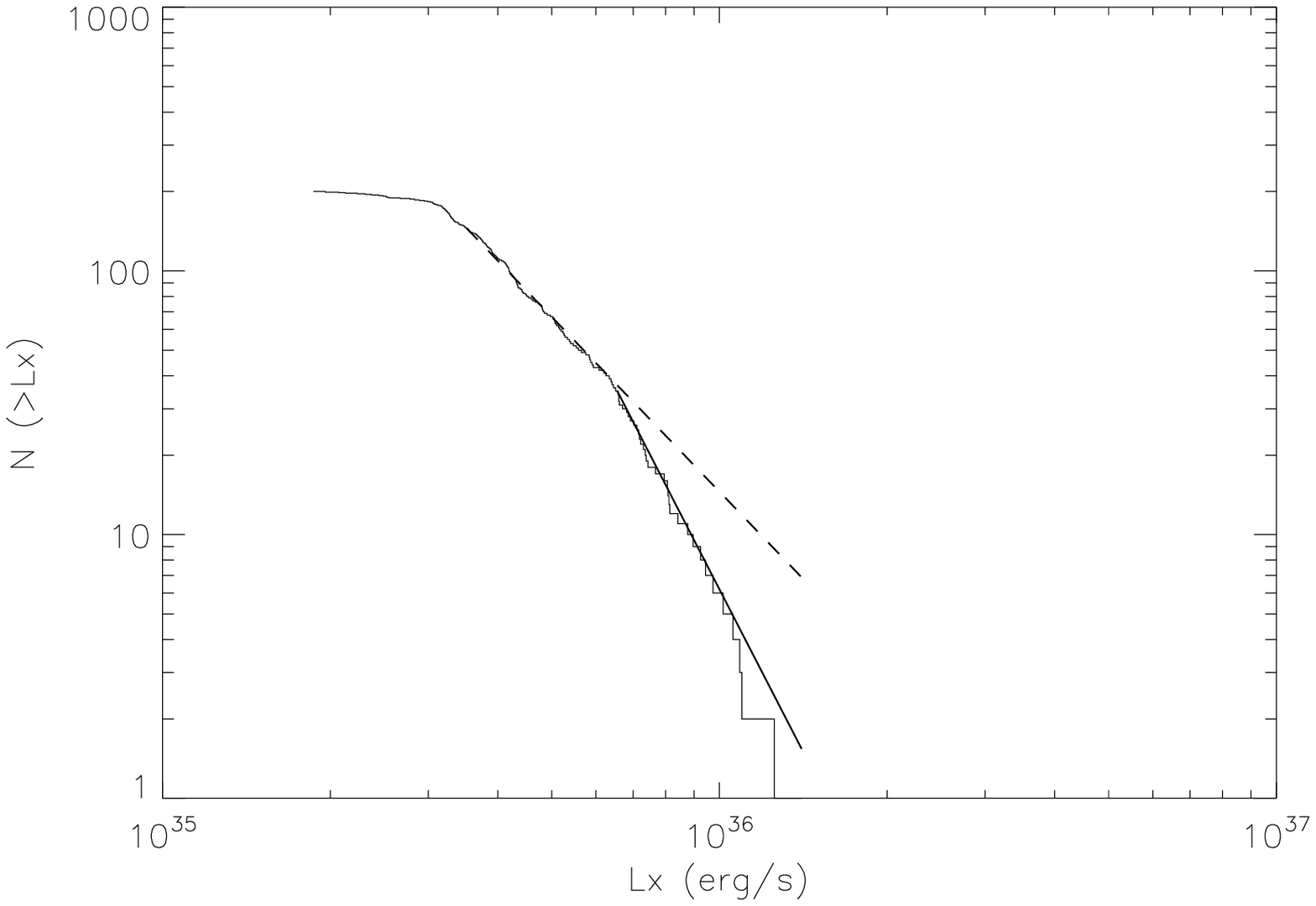} & 
\includegraphics[height=5.8cm, angle=0]{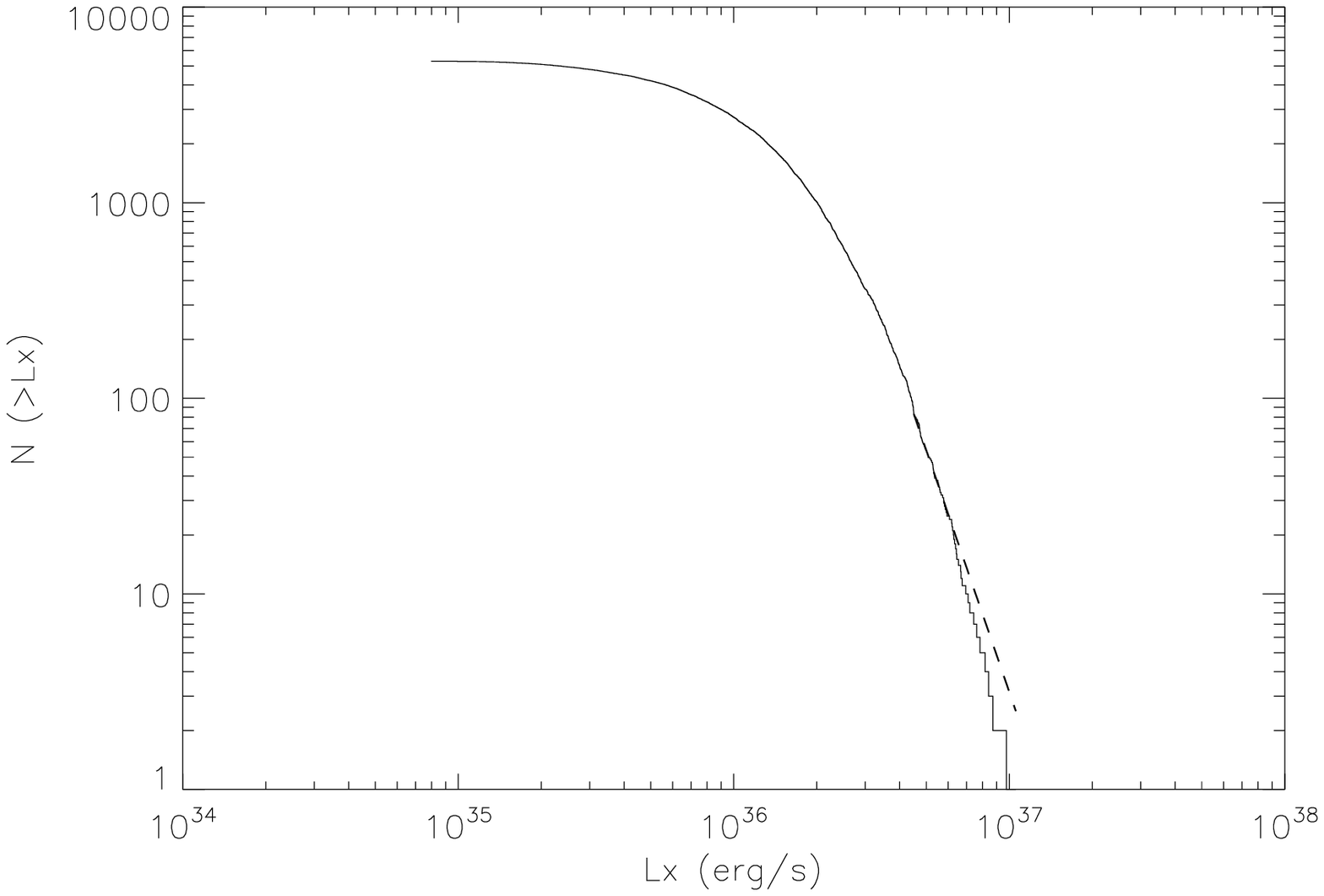}
\end{tabular}
\caption{Examples of  power-law fits (dashed lines) to the non-normalized cumulative luminosity distributions (17--50\,keV; solid lines) of four sources: two prototypical SFXTs (SAX~J1818.6--1703, {\em upper left}; 
IGR~J17544--2619, {\em upper right}), the intermediate SFXT IGR~J18483--0311 ({\em lower left}, with two power-laws, before and after a break) and the classical HMXB 4U~1700--377  ({\em lower right} panel), where only the tail is power-law like.}
\label{lsfig:lcurves}
\end{figure*}

Figure~\ref{lsfig:lcurves} shows some examples of power-law fits to the cumulative distributions for the three typical cases: 
single power-law fit for most of the data (after the truncation point, upper panels), two power-law fit for most of the data (after the truncation point, lower left panel) and single power-law fit for a very limited fraction of the data (the high energy tail, lower right panel).

\begin{figure*}
\begin{center}
\centerline{\includegraphics[width=14cm,angle=0]{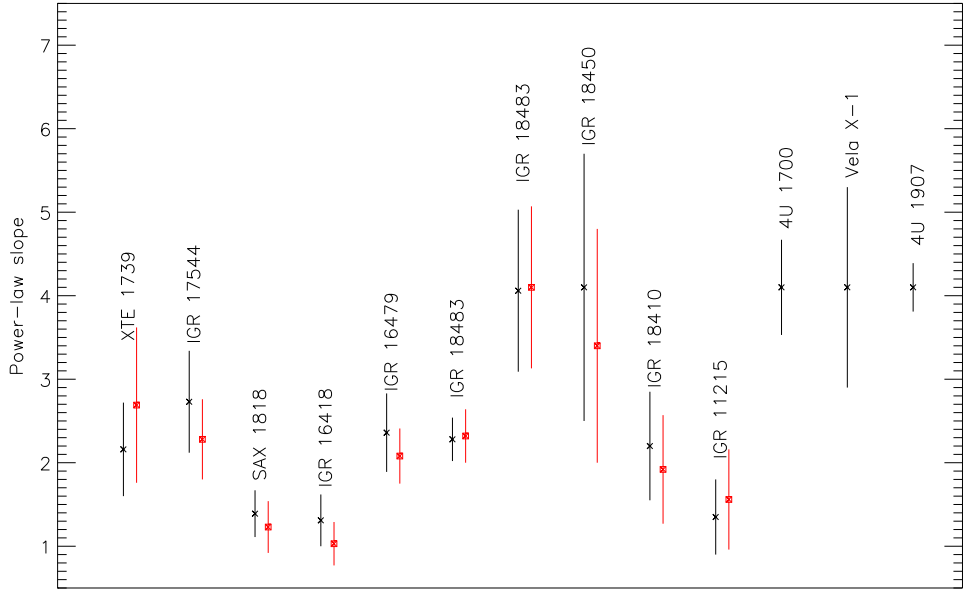}} 
\caption{Results of the MLE of the power-law  parameters of the cumulative luminosity distributions 
in the energy band 17--50 keV ({\em crosses}) and in the 22--50 keV ({\em squares}), for comparison. }
\label{fig:slopes}
\end{center}
\end{figure*}

In Figure~\ref{fig:slopes} we show the power-law slopes that better model the
cumulative luminosity distribution functions in the energy range 17--50 keV (black {\em crosses})
and in the energy band 22--50 keV (red {\em squares}). 
For classical HMXBs we have only the results in the 17--50\,keV range.
A very good agreement is found between the power-law slopes estimated
in the two IBIS/ISGRI energy bands. Hence the effect of the instrumental LT 
fluctuations up to 22\,keV  plays a role in the percentage of detections alone, with no further 
bias involved.

We note also that in case of SFXTs, a good fit with a power-law is obtained considering
a large subsample ($>$80\%) of the original data points (SAX~J1818.6--1703, IGR~J18410--0535
and IGR~J11215--5952; in this latter source 100\% of the data points follow a power-law model).
The classical HMXBs however, behave very differently: a 
power-law model is a good fit only for the high luminosity tail of the distribution
(involving less than 0.5\%, 1.6\% and 20\% of the total data points for Vela~X--1, 4U~1700--377
and 4U~1907$+$09, respectively). We note that their power-law slope is always around 4.
Apart from  IGR~J18450--0435, for which the sample is small and the uncertainty on the power-law slope is large,
the source IGR~J18483--0311 also displays a behavior which shares some similarity with the classical HMXBs:
its cumulative distribution appears to show a break at about 6.7$\times$10$^{35}$ erg~s$^{-1}$, with
a slope of 2.28$\pm{0.26}$ before it (more in line with the flatter power-laws shown by other SFXTs)
and a slope of 4.06$\pm{0.97}$ above it (compatible with the tail in the classical HMXBs).
This quantitatively confirms the intermediate character of this SFXT \citep{RahouiChaty2008}.

We are aware that for some sources the sample used for MLE of the 
power-law exponent is small, and that a high KS probability does not
imply that the data are truly drawn from a power-law distribution.
However, these quantitative estimations confirm and  better quantify what was already evident
from the visual inspection of the cumulative distributions: SFXTs are more power-law like (and with a flatter slope)
in their cumulative luminosity distributions with respect to the classical HMXBs, with SAX~J1818.6--1703 and IGR~J12215--5952 
being the best cases of power-law like distribution.

\begin{table*}
 \centering
  \caption{Maximum Likelihood Estimation of the power-law  parameters of the cumulative luminosity distributions (17--50\,keV) }
  \begin{tabular}{@{}lcccc@{}}
\hline
   Name             &  Truncation point                 & \#ScWs (tot \#ScWs)&  Power-law slope      &   KS probability  \\
                    &  L$_{X}$ (10$^{35}$ erg~s$^{-1}$) &                  &                 &            \\
\hline
Prototypical SFXTs  &                                   &                  &                 &            \\
\hline
XTE~J1739--302      &      6.3                          &       32  (87)   & 2.16$\pm{0.56}$ &   0.984    \\
IGR~J17544--2619    &      6.5                          &       31  (64)   & 2.73$\pm{0.61}$ &   0.953    \\
SAX~J1818.6--1703   &      1.4                          &       41  (46)   & 1.39$\pm{0.28}$ &   0.993    \\
\hline
Intermediate  SFXTs &                                   &                  &                 &            \\
\hline
IGR~J16418--4532    &     52.0                          &       36  (57)   & 1.31$\pm{0.31}$ &  0.997     \\ 
IGR~J16479--4514    &     14.6                          &       52 (172)   & 2.36$\pm{0.47}$ &  0.996     \\
IGR~J18483--0311    &      3.6                          &      139 (200)   & 2.28$\pm{0.26}$ &  0.962     \\
                    &      6.7                          &       30 (200)   & 4.06$\pm{0.97}$ &  0.979     \\
IGR~J18450--0435    &      4.48                         &       15  (16)   & 4.1$\pm{1.6}$   &  0.861     \\
\hline
Less explored SFXTs &                                   &                  &                 &            \\
\hline
IGR~J18410--0535    &      3.8                          &       24  (29)   & 2.20$\pm{0.65}$&  0.899      \\
\hline
The periodic SFXT   &                                   &                  &                 &             \\
\hline
IGR~J11215--5952    &      $-$                          &       19  (19)   & 1.35$\pm{0.45}$&  0.885     \\
\hline
\hline
HMXBs               &                                   &                  &                 &           \\
\hline
4U~1700--377        &      45.0                         &       83 (5285)  & 4.10$\pm{0.57}$ &  0.891     \\
Vela~X--1           &      45.0                         &       10 (1970)  & 4.1$\pm{1.2}$   &  0.958     \\
4U~1907$+$09           &       6.56                        &      210 (815)   & 4.10$\pm{0.29}$ &  0.721      \\
\hline
\end{tabular}
\label{tab:powfit}
\end{table*}

\begin{figure}
\begin{center}
\begin{tabular}{cc}
\includegraphics[width=8.cm,angle=0]{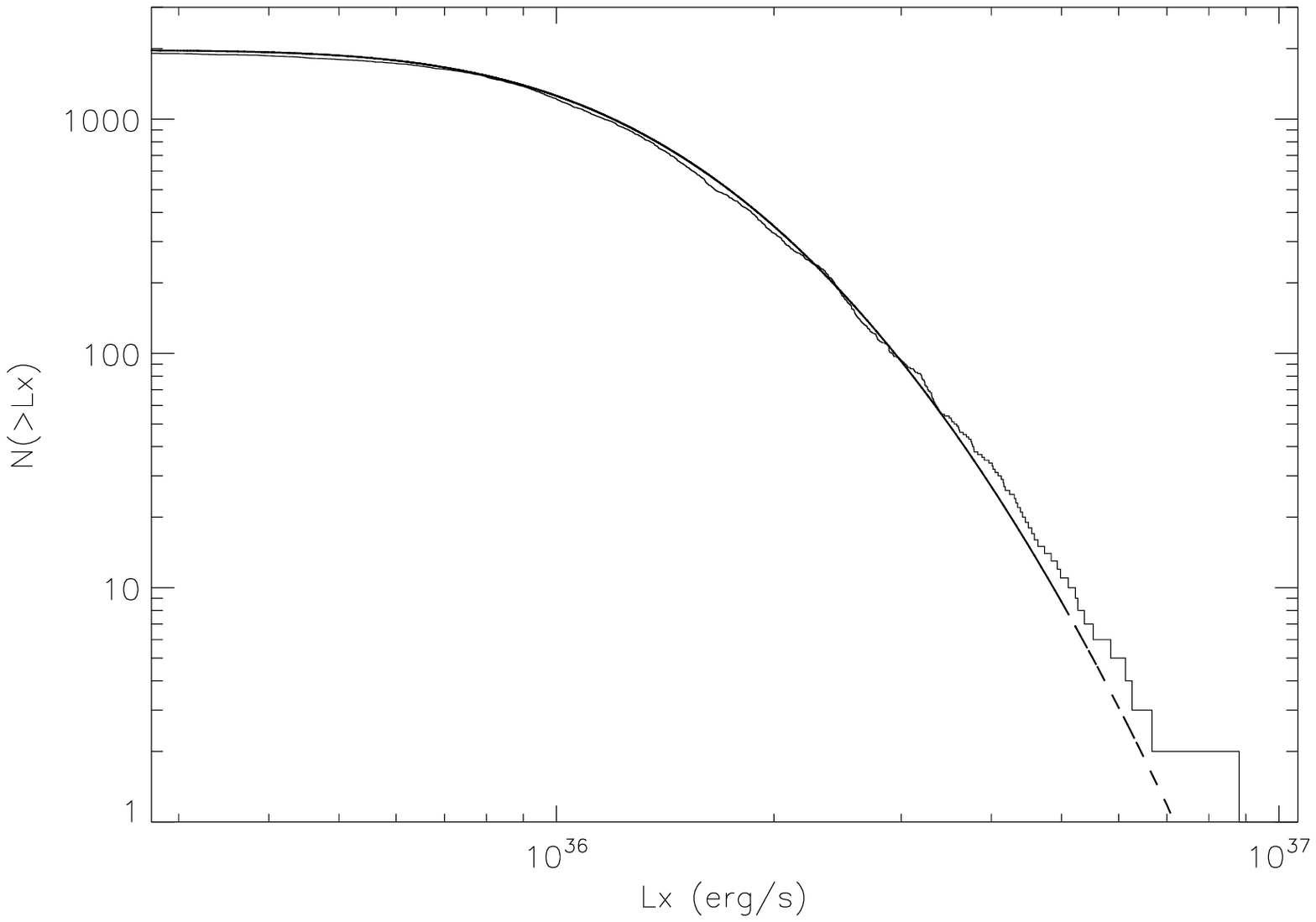} \\
\includegraphics[width=8.cm,angle=0]{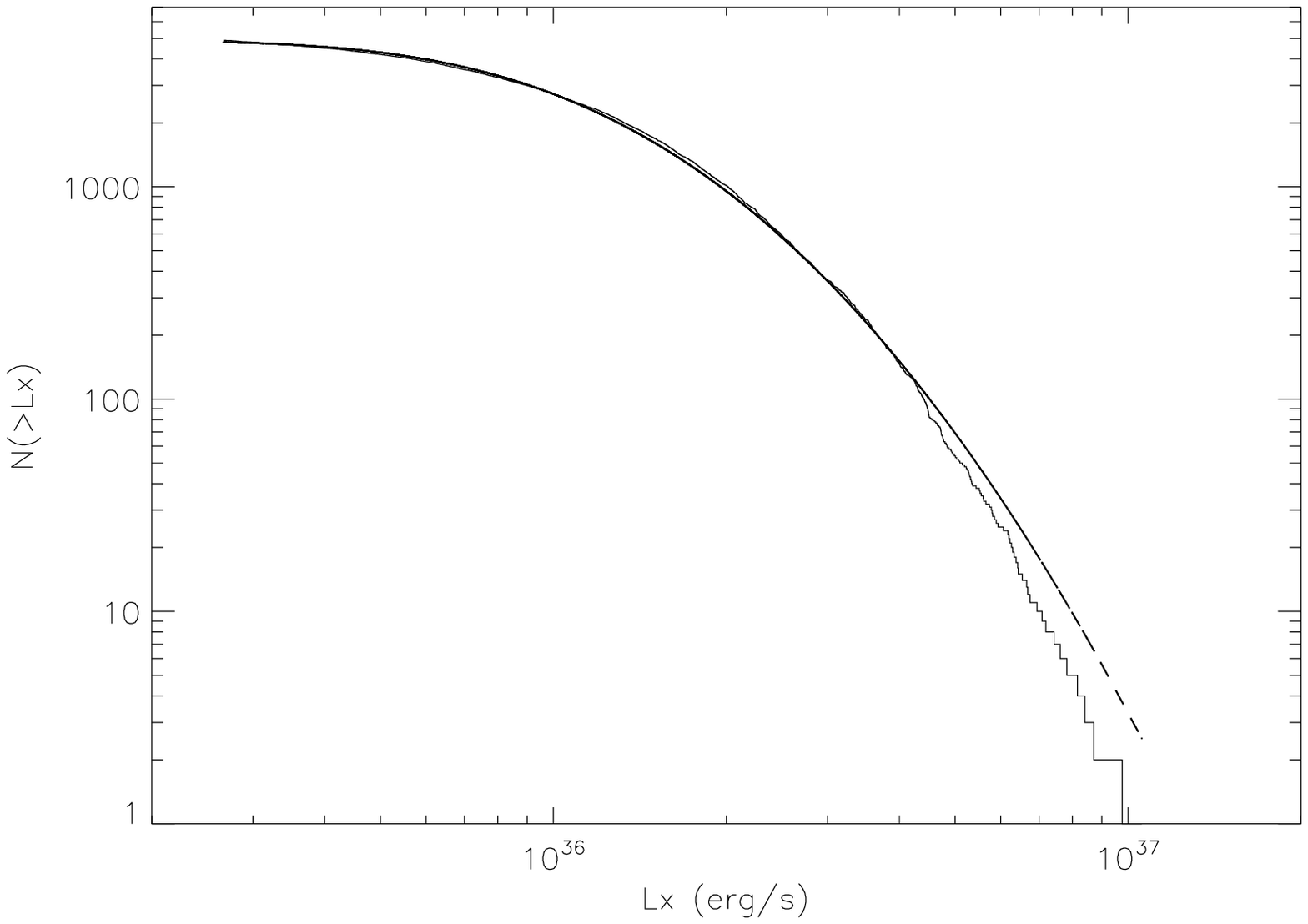}
\end{tabular}
\end{center}
\caption{Examples of the best log-normal functions (dashed lines) fitted to the Vela~X--1  ({\em upper panel}) 
and 4U~1700--377  ({\em lower panel}) non-normalized cumulative luminosity 
distributions (solid lines; 17--50 keV).  See text for the details and the resulting parameters.} 
\label{fig:lognormfit}
\end{figure}

Since the luminosity distributions of 4U~1700--377 and Vela~X--1 are completely different from those
shown by other sources, and since the power-law fit in their case can account for only the high luminosity tail,
we tried to fit their 17--50\,keV curved shape with the cumulative distribution expected from a log-normal function.
Indeed, a log-normal function was found to  fit well the X--ray brightness distribution of Vela~X--1 by \citealt{Furst2010}, using \inte~data (20--60\,keV). 
In our case, MLE of the parameters of the cumulative distribution for a log-normal function resulted in 
a median luminosity of 1.2$\times$10$^{36}$~erg~s$^{-1}$    and  1.1$\times$10$^{36}$~erg~s$^{-1}$  
(with a multiplicative standard deviation, $\sigma$, of 0.542  and  0.688)
for Vela~X--1 and 4U~1700--377, respectively (Figure~\ref{fig:lognormfit}).
The best truncation points for the two cases were 2.75$\times$10$^{35}$~erg~s$^{-1}$ (Vela~X--1, implying 1909 out of 1970 data points) and 2.67$\times$10$^{35}$~erg~s$^{-1}$ (4U~1700-377, implying 4903 out  of 5285 data points). 
Note however that in our case a log-normal can be rejected (KS probability of 0.408 and 0.0225 for Vela X--1 and 4U~1700-377, respectively).
Indeed, although a log-normal may be a plausible representation of data spanning a short period of time \citep[as in][ \inte~revolutions 0433--0440]{Furst2010}, this is not the case when a wider database is considered (as in this work, where Vela~X--1 detections spanning revolutions 0028--1138 are considered). This can be explained
by the presence of long-term trends in nine years of \inte~data, trends that are clearly visible in the case of persistent and bright sources (such as Vela X--1 or the already discussed Crab), but  buried within the highly variable emission and more sporadic detections of  SFXTs.

We will not discuss this further, since the detailed investigation of 
the exact shape of the cumulative distributions in these 
two persistent sources is beyond the scope of our paper.

	\section{Discussion} \label{sec:discussion}

IBIS/ISGRI is able to catch bright SFXT flares that represent
the most extreme events in HMXBs hosting neutron stars.
The temporal profiles of SFXT flares are usually very complex (e.g., \citealt{Sidoli2011texas})
so it is problematic to define an {\em X--ray flare} given also the
ambiguity in clearly separating overlapping single flares.
Since time scales of flare durations are roughly consistent with a ScW 
duration, we decided to consider the single IBIS/ISGRI detections on ScW level (IMA 
results; $\sim$\,ks sampling) as a good representation of SFXT flares.
SFXT flares are intermittent with a duration much shorter than the time
interval between two adjacent major outbursts (typically a few months).
An outburst is composed of several flares with different
peak luminosity, duration, and temporal profile. Their occurrence cannot be a priori predicted,
except for the periodic SFXT IGR~J11215--5952 \citep{SidoliPM2006}, where the periodically recurrent
outbursts are believed to trace the orbital period of the system.

Using the whole \inte~archive available to date (spanning about nine years),
we have built the cumulative distributions of the luminosity of all known SFXTs,
and compared them with classical HMXBs. 
For the first time, we have quantitatively characterized such distributions for the SFXTs, 
deriving their duty cycles, dynamic range, and typical luminosity in bright flaring activity.
Indeed, the cumulative distributions show many aspects of the
behaviour of sources at hard X--rays, when comparing SFXTs to the other three HMXBs.
In Figure~\ref{fig:ima3} we have shown the cumulative luminosity distributions at \ima~level in the 
energy range 17--50\,keV, which is the most comprehensive (and with the best statistics) at \ima~level.
On the x-axis the range of variability (the dynamic range as observed by IBIS/ISGRI)
covered by each source is evident, together with the threshold in X--ray luminosity above which
\inte~can observe each source, assuming that the distances are correct. 
On the y-axis, the percentage of time spent above the threshold of detectability 
by the different sources can be derived.
Clearly different behaviors can be seen from these cumulative distributions.
Vela~X--1 and 4U~1700--377 are always detected except than during X--ray eclipses or 
the so-called \lq\lq off-states\rq\rq~\citep{Kreykenbohm2008}.
Their most frequent state is at high luminosity, around 1-2$\times$10$^{36}$ \ergsec.
The transient HMXB 4U~1907$+$09 distribution is located below them, implying a duty cycle
of $\sim$20\%,  intermediate between the persistent HMXBs (Vela~X--1 and 4U~1700--377) 
and the {\em intermediate} SFXTs
IGR~J18483--0311 and IGR~J16479--4514 (with a duty cycle of $\sim$4-5\%).
This confirms the (not so) transient behavior previously reported
for this source \citep{Doroshenko2012}.

Orbital and super-orbital modulations of the source X-ray light curves could, in principle (as any
other periodic or aperiodic trend), affect the shape of the cumulative luminosity distributions.
However, considering both the nature of the sources in our study and the {\it INTEGRAL}
sensitivity (able to catch only the bright flares), the effect of the amplitude of the orbital and
super-orbital modulations is negligible when compared to the X-ray luminosity variability produced
by the SFXTs X-ray flares. A similar conclusion holds also for the cumulative luminosity
distributions of persistent HMXBs, but for a different reason: here the orbital modulations are
due to the presence of X--ray eclipses, during which the sources are not detected with {\it INTEGRAL},
so that the eclipses do not produce any effect in the cumulative distributions.

The sources with the lowest \ima~based duty cycle are confirmed to be the {\em prototypical SFXTs} together
with some less explored SFXTs and the periodic SFXT IGR~J11215--5952.\footnote{We note that the distance to IGR~J16418--4532 is not very constrained ($\sim$13 kpc), so
its position along the x-axis is not well known, and could be shifted to lower luminosities.}
However, when we consider the short-term variability (100\,s, Table~\ref{tab:lcr_3s}) we note that the prototypical SFXTs are detected more than 20\% of the times, while the intermediate ones reach about 14\% at most (17\% at most for the less explored SFXTs). This result holds also if we increase the \lcr~detection threshold to the more conservative 5$\sigma$ level, with the  the prototypical SFXTs being around 8\% and the remaining (intermediate and less explored) SFXTs being below 1\% (the only exception being IGR~J08408--4503 that reaches 6\%). 
This result does not depend on the number of detections in the \ima~step (i.e. more detections in \ima~resulting in more detections in \lcr), since, for example, IGRJ~16479--4514 has 111 detections at a ScW level (Table~\ref{res:ima}) and an 11\% of detections in \lcr~(Table~\ref{tab:lcr_3s}), while XTE~J1739--302 has  only 70 detections at a ScW level and about 27\% of detections in \lcr. 
Furthermore, this difference is unlikely due to statistics alone  (i.e. the brightest sources having the highest detection percentage in 100\,s) since, as can be seen from Table~\ref{res:ima}, there is not a clear and linear behaviour in the sources, e.g. 
IGR~J18410--0535 has an \ima~average count-rate of 8$\pm$0.3\,\countsec~and an \lcr~detection of 17\%, while IGR~J16418--4532 has an \ima~average rate of 9$\pm$0.3\,\countsec~and an \lcr~detection of 7\%.
Hence it is likely that we are seeing a real difference within the sub-groups of SFXTs: when investigating how the 100\,s bins are distributed within the \lq\lq good \rq\rq~ScWs, we find that the intermediate SFXTs seem to be more constant, with most weak bins giving the detection in a ScW, while the prototypical ones are more variable, with more flaring bins that lead to the detection at a ScW level.  

The cumulative distributions of most of the SFXTs analyzed here can plausibly be 
considered to follow a power-law function.
The power-law slope of the cumulative distributions for the three prototypical SFXTs
also agrees with what previously found with PCA/RXTE data \citep{Smith2012}.
Although Smith and collaborators  perform only a rough estimate of the power-law
(slope of 1.5, with no uncertainty), 
the slopes we found for the three prototypical SFXTs at hard X--rays with \inte~seem to be consistent with the power-law like luminosity distribution derived at soft X--rays (2--10\,keV).
Although the X--ray luminosity range covered by SFXTs in our study is about 
two orders of magnitude, being the luminous outbursts covered,
a plausible fit with a power-law 
of their cumulative luminosity distributions is reminiscent of Self-Organized Criticality (SOC) Systems 
(e.g. \citealt{Aschwanden2013} and references therein).
A SOC \citep{Bak1987} is a system which naturally and perpetually evolves into a critical state
where a minor event can start a chain reaction leading to a catastrophe, 
like in a sandpile, where even though sand is uniformly added to the pile, 
the amount of sand falling from the pile can 
greatly vary with time, giving rise to unpredictable ``avalanches'' when a certain instability threshold is reached.
Several phenomena, not only in nature but also in human systems, are believed to behave like SOC systems: 
earthquakes, landslides, solar flares, forest fires, lunar craters (e.g. \citealt{Newman2005}).

The power-law scaling of SFXTs cumulative luminosity distributions, a necessary but not sufficient condition
for a SOC system, is remarkable.
We suggest that SFXTs flares can be possibly considered as  ``avalanches'' in SOC systems, which
are triggered when a critical state is reached.
In our case, accretion can be interpreted as the slow and steady driver towards the critical state required to the SOC system
to produce the avalanche.

Recently, \citet{Shakura2012} proposed a new model to explain quasi-spherical accretion
of matter in HMXBs with neutron stars (NS) and supergiant companions, and it has been successfully applied
to SFXTs by \cite{Drave14}.
In this model, if the source is in a low luminosity state ($<$3-5$\times$10$^{35}$\ergsec),
the wind matter captured by the NS within the Bondi radius cannot efficiently cool
down  by Compton processes; a hot shell forms between the Bondi radius and the magnetospheric radius,  accumulating
above the NS magnetosphere.
This matter penetrates the magnetosphere {\em only} if it is able to cool down 
to a critical temperature, so that Rayleigh-Taylor instability can allow high accretion rates onto the NS.
When this critical temperature is reached, if a considerable amount of matter has already accumulated in the hot shell,
it can suddenly accrete  onto the NS producing SFXTs outbursts.
In this respect, we suggest that this critical temperature can act as the threshold needed to start
the intermittent \lq\lq avalanche\rq\rq~in a SOC system.
This is very different from the persistent and bright HMXBs case, where a high photon flux allows the matter to efficiently
cool down thanks to Compton cooling, maintaining high accretion rates.

	\section{Conclusion \label{concl}}

We used about nine years of publicly available  \inte~observations to study in a systematic way 
all known SFXTs and three classical HMXBs.
We have built the cumulative luminosity distributions and, for the first time for SFXTs, 
we have quantitatively characterized them. We  derived duty cycles, dynamic ranges, slow versus fast variability properties and  typical luminosities during bright flaring activity.

We characterized the cumulative luminosity distributions with a power-law model.
We found that the X--ray luminosity range where the model is a plausible representation of the data is larger
for SFXTs than for persistent HMXBs (where a power-law is able to reproduce
only the very high luminosity tail). Furthermore,
SFXTs show a significantly flatter power-law slope than HMXBs.

We suggest that this power-law like behaviour is a possible indication of self-organized criticality and
that SFXTs flares could be associated with avalanching resulting in a SOC system
when an instability threshold is reached.

Among the several different explanations suggested for SFXTs, the \cite{Shakura2012} model
for quasi-spherical accretion seems, to date, to better explain SFXTs and their link with persistent
systems \citep[see][for the application to SFXTs]{Drave14}.
It predicts the complete collapse of a hot shell of 
gravitationally captured wind material
accumulated above the NS magnetosphere, when it cools below the critical temperature,
allowing the onset of Rayleigh-Taylor instability.

While persistent accreting pulsars spend most of their life in an efficient Compton cooling
regime which allows (and is permitted by) high X--ray luminosities (10$^{36}$\ergsec), 
SFXTs spend most of the time at much lower luminosities (10$^{33}$--10$^{34}$\ergsec; \citealt{Sidoli2008:sfxts_paperI})
where gravitationally captured matter accumulates above the NS magnetosphere and remains too hot
to accrete at high rates. At these low X--ray luminosities, the matter is subject only to inefficient
radiative cooling which allows a low accretion rate through the NS magnetosphere.
At some point, the X--ray luminosity produced by this inefficient accretion can reach 
the critical X--ray luminosity which is able to trigger efficient Compton cooling of the
matter below the critical temperature.
This produces the onset of Rayleigh-Taylor instabilities and the complete collapse 
of the accumulated hot shell, giving rise to a SFXT outburst.

To conclude, we suggest that SFXTs power-law like cumulative luminosity distributions are possibly the consequence
of a threshold-nature instability, supporting the \cite{Shakura2012} model for quasi-spherical settling accretion regime.

\section*{Acknowledgments}

%
Based on observations with \textit{INTEGRAL}, an ESA project
with instruments and science data centre funded by ESA member states
(especially the PI countries: Denmark, France, Germany, Italy,
Spain, and Switzerland), Czech Republic and Poland, and with the
participation of Russia and the USA. 
This work has made use of the \inte~archive developed at INAF-IASF Milano, 
http://www.iasf-milano.inaf.it/$\sim$ada/GOLIA.html.
 We acknowledge the Italian Space Agency financial  support {\it INTEGRAL}  
 ASI/INAF agreement n. 2013-025.R.0.
AP and LS thank Tony Bird, Massimo Cocchi, Mariateresa Fiocchi, Aleksandra Gros, Sandro Mereghetti, 
Lorenzo Natalucci and Jerome Rodriguez for useful discussions.

\bsp

\label{lastpage}

\end{document}